\documentclass[lettersize,journal,dvipsnames]{IEEEtran}

\usepackage{amsmath,amsfonts}
\usepackage{algorithmic}
\usepackage{algorithm}
\usepackage{array}
\usepackage[caption=false,font=normalsize,labelfont=sf,textfont=sf]{subfig}
\usepackage{textcomp}
\usepackage{stfloats}
\usepackage{url}
\usepackage{verbatim}
\usepackage{graphicx}
\usepackage{cite}
\usepackage{tabularray}
\usepackage{rotating}
\usepackage{pdflscape}
\usepackage[most]{tcolorbox}
\usepackage{lipsum} 
\usepackage{etoolbox}
\usepackage{adjustbox}
\usepackage{tikz}
\usepackage{orcidlink}
\usepackage{amssymb}
\usepackage{multirow}
\usepackage{booktabs}
\usepackage{mdframed}

\DeclareFontFamily{U}{stix2bb}{}
\DeclareFontShape{U}{stix2bb}{m}{n} {<-> stix2-mathbb}{}

\NewDocumentCommand{\indicator}{}{\text{\usefont{U}{stix2bb}{m}{n}1}}

\definecolor{tab10blue}{rgb}{0.09019607843137255, 0.7450980392156863, 0.8117647058823529}  
\definecolor{tab10orange}{rgb}{1.0, 0.4980392156862745, 0.054901960784313725}  
\definecolor{tab10green}{rgb}{0.17254901960784313, 0.6274509803921569, 0.17254901960784313}  
\definecolor{tab10red}{rgb}{0.8392156862745098, 0.15294117647058825, 0.1568627450980392}  
\definecolor{tab10purple}{rgb}{0.5803921568627451, 0.403921568627451, 0.7411764705882353}  
\definecolor{black}{rgb}{0.0, 0.0, 0.0}  

\usepackage{subfiles}
\usepackage{hyperref}

\newcommand{\rev}[1]{\textcolor{black}{#1}}

\PassOptionsToPackage{svgnames,x11names,dvipsnames}{xcolor}
\usepackage[most]{tcolorbox}

\newtcbox{\inlinebox}[1][]{enhanced,
 box align=base,
 nobeforeafter,
 colback=cyan,
 colframe=cyan,
 size=small,
 left=0pt,
 right=0pt,
 boxsep=1pt,
 fontupper=\footnotesize,
 #1}

\begin{document}

\title{\rev{SELU: A Software Engineering Language Understanding Benchmark}}

\author{Fabian C. Peña \orcidlink{0009-0008-2249-7990}, Steffen Herbold \orcidlink{0000-0001-9765-2803}
\thanks{This work is funded by the German Research Foundation (DFG) through the SENLP project under Grant 524228075.}
\thanks{The authors are with the Chair of AI Engineering, University of Passau, Passau 94032, Germany (e-mail: fabiancamilo.penalozano@uni-passau.de; steffen.herbold@uni-passau.de).}
}



\maketitle

\begin{abstract}
\rev{Large Language Models (LLMs) have demonstrated remarkable capabilities in code understanding and generation. However, their effectiveness on non‐code Software Engineering (SE) tasks remains underexplored. We present `Software Engineering Language Understanding' (SELU), the first comprehensive benchmark for evaluating LLMs on 22 SE textual artifacts NLU tasks, spanning from identifying whether a requirement is functional or non-functional to estimating the effort required to implement a development task. SELU covers classification, regression, Named Entity Recognition (NER), and Masked Language Modeling (MLM) tasks, with data drawn from diverse sources such as issue tracking systems and developer forums. We fine-tune 22 open-source LLMs, both generalist and domain-adapted; and prompt two proprietary alternatives using zero-shot a 3-shot prompting strategies. Performance is measured using metrics such as F1-macro, SMAPE, F1-micro, and accuracy, and compared via the Bayesian signed-rank test. Our results show that fine-tuned models across various sizes and architectures perform best, exhibiting high mean performance and low across-task variance. Furthermore, domain adaptation via code-focused pre-training does not yield significant improvements and might even be counterproductive for developer communication tasks.}
\end{abstract}


\section{Introduction}

\IEEEPARstart{L}{arge} Language Models (LLMs) have become increasingly popular in many industries, including Software Engineering (SE). Through tools such as GitHub Copilot \cite{copilot} and Cursor \cite{cursor}, developers are able to integrate LLMs into daily tasks, taking advantage of their still evolving capabilities to understand and generate code. Although these tools can enable an increase in efficiency in specific tasks such as documentation and code completion, they still struggle with more complex tasks, large functions, multiple files, and proprietary contexts \cite{copilot-survey}. Moreover, the typical SE life cycle encompasses more than just coding, and additional activities related to requirements analysis, software design, quality assurance, software maintenance, and software management are also critical to ensure the success of a real software product over time. 

In recent years, the academic community has put a great effort into how LLMs can be applied throughout the SE life cycle; however, despite the wide variety of works, most of them focus on code-related tasks (e.g., code generation and code understanding) \cite{llm4se1,llm4se2,llm4se3,llm4se4,llm4se5}. Consistently, widely adopted SE-specific benchmarks mainly aim to evaluate LLM capabilities in code generation \cite{humaneval,mbpp,evalplus,swebench,bigcodebench,evoeval,codeelo,fauneval}, lacking diversity of data sources and task types \cite{benchmarks-criticism}.

\IEEEpubidadjcol

To the best of our knowledge, in addition to these individual efforts to evaluate LLMs on specific tasks, there are no works intended on setting the foundations to evaluate their capabilities from a broader perspective in the SE domain, particularly on \rev{SE textual artifacts-related} tasks. With a broader perspective, we refer to two interrelated aspects: (1) Considering multiple tasks and task types, i.e., tasks belonging to activities throughout the SE life cycle, with data coming from a variety of sources and different types of targets. (2) Establishing a more robust evaluation approach that extends from simple performance score aggregations to, e.g., pairwise model comparisons via a Bayesian analysis.

To address the first aspect, \rev{inspired by GLUE/SuperGLUE~\cite{glue,superglue}}, we create a new \rev{curated} benchmark by collecting from the literature \rev{22 SE textual artifacts Natural Language Understanding (NLU) tasks} of type classification, regression, Named Entity Recognition (NER), and Masked Language Modeling (MLM). \rev{They are built from sources such as issue tracking systems, developer forums, requirements specifications, among others; and, for a more context-aware fine-grained analysis, we group them in two different taxonomies: SE life cycle activity (requirements analysis, software development, quality assurance, software maintenance, and software management tasks) and task family (code-adjacent metadata, core non-code, and developer communication tasks). Some of them include: (i) classifying a requirement as functional or non-functional, (ii) estimating the effort required to implement a development task, and (iii) identifying SE-specific terminology discussed in a post.} We name this benchmark `Software Engineering Language Understanding' (SELU).

For the second aspect, we fine-tune 22 open-source LLMs and prompt two proprietary alternatives on \rev{the tasks included in SELU} to answer the research question: \rev{\textit{Do the selected LLMs perform consistently across different SE textual artifacts NLU tasks?}} \rev{To answering this question, we first apply pairwise model comparisons via the Bayesian signed-rank test. Then,} to better understand the factors driving model performance, we analyze the impact of performance drivers such as model size, architecture, and domain adaptation via code-focused pre-training\rev{, as well as task family and task type}. \rev{Our methodological contributions also include a unified size-aware fine-tuning to better control the learning process on small datasets; standardized zero-shot and 3-shot prompting strategies to enable fairer comparisons between proprietary and open-source LLMs; and the inclusion of per-task baseline models trained using two classical machine learning (ML) approaches} to evaluate the performance improvement that LLMs can offer to compensate for their increasing complexity. \rev{Fully generation and software design-related tasks are out-of-scope in this version of SELU}.


The paper is organized as follows. Section~\ref{s:background} reviews the background and related work on LLMs in SE. In Section~\ref{s:benchmark}, we describe how we collect the tasks and prepare the datasets. Section~\ref{s:methodology} presents the models, implementation details, evaluation procedure, and Section~\ref{s:results} highlights the most important findings. In Section~\ref{s:discussion}, we discuss these findings, derive implications for practitioners, and outline future work, taking into account the threats defined in Section~\ref{s:threats}. Finally, Section~\ref{s:conclusions} summarizes our main contributions. All source code, datasets, prompt templates, and results are available in our replication package~\cite{replicationpackage}.

\section{Background and Related Work}
\label{s:background}

In this section, we survey existing research on LLMs in SE, first by examining how LLMs have been integrated into code-centric workflows and the benchmarks developed to evaluate their coding capabilities. We then broaden the lens to review studies applying LLMs to non-code tasks throughout the SE life cycle, highlighting the gaps in current evaluations that our work seeks to address.

\subsection{The Mainstream of LLMs in SE}

LLMs are rapidly becoming an integral part of SE, where recent models such as Claude 3.7 \cite{claude37} and GPT-4o \cite{gpt4o} can understand and generate code from natural language prompts. These LLMs are the core of a new generation of integrated development environments (IDEs) such as GitHub Copilot \cite{copilot} and Cursor \cite{cursor}, which offer: (1) context-aware code completion and generation, (2) intelligent code refactoring and documentation assistance, and (3) seamless integration with version control, debugging tools, and team collaboration workflows. These tools can reduce developer effort by about 35\% and deliver up to 50\% time savings on routine tasks such as writing comments, generating documentation, and automating CI/CD workflows; with the biggest boost in languages such as JavaScript, Java, Go, and Python, whilst often producing lower quality or incorrect code for performance‐critical C and C++ components \cite{copilot-survey}. They can also produce clear explanations of code and help with creating unit tests and system monitors, but struggle to modify large, multi‐file proprietary codebases without significant developer oversight.

To gain some insight into how LLMs perform in code-related tasks, early benchmarks such as HumanEval \cite{humaneval} and MBPP \cite{mbpp} evaluate LLMs using synthetic, function-level Python tasks designed to assess basic coding capabilities. Subsequently, EvalPlus \cite{evalplus} extends HumanEval and MBPP with adversarial test cases to better measure functional correctness. Faced with the need to evaluate LLMs in more realistic scenarios, SWE-bench \cite{swebench} introduces resolution of real-world GitHub issues. Later, BigCodeBench \cite{bigcodebench} and EvoEval \cite{evoeval} emphasize a broader coverage and evolving task difficulty, while CodeElo \cite{codeelo} proposes a competition-style ranking to better capture model capability levels beyond pass rates.

Despite these efforts, SE benchmarks still lack in diversity of sources and task types, in addition to the need to include evaluation metrics that reflect model usability \cite{benchmarks-criticism}. To the best of our knowledge, the only benchmark that goes further in this direction is FAUN-Eval \cite{fauneval}, which decomposes issue resolution into three sub-tasks (question-answering, fault localization, and code editing), providing fine-grained insights into LLM performance across different phases of bug fixing.

\subsection{\rev{NLP Throughout the SE Life Cycle}}

Even with the importance of coding, a substantial portion of SE work involves non-code \rev{(or rather textual artifacts-related)} tasks, which span the entire SE life cycle. Requirements analysis, software design, quality assurance, software maintenance, and software management are also critical activities to ensure the success of a real software product over time. Previous studies report that developers often spend only about 15\% of the time reading and writing code; and another 14\% debugging and fixing bugs \cite{productivity1,productivity2}. The rest of the time is spent on running tests, defining specifications, reviewing code, documenting, joining meetings, among others.

In recent years, the academic community has put a great effort into how LLMs can be applied throughout the SE life cycle; even so, Hou et al.~\cite{llm4se2} report that of 395 works surveyed, about 79\% address the application of LLMs only in software development and maintenance activities, underscoring a strong focus on code-related tasks. Zhang et al. \cite{llm4se5} highlight the urgent need for domain-specific LLMs and more usability-oriented benchmarks that go beyond traditional code-focused evaluations. Finally, Fan et al. \cite{llm4se1} emphasize that hallucinations continue to challenge the reliability of LLM-generated artifacts and advocate combining LLMs with automated regression oracles and hybrid testing strategies to mitigate erroneous outputs.

\rev{Before LLMs emerged, other Natural Language Processing (NLP) approaches were applied to different SE tasks with partial success. For instance, parsing-based tools such as NEON~\cite{neon} help in the automatic extraction of patterns/rules that can be validated by a human-in-the-loop to support tasks like app reviews classification, while Kallis et al.~\cite{fasttext-github} demonstrate that FastText~\cite{fasttext} (a library for learning text representations) is able to produce an acceptable performance in tasks such as GitHub issues classification. Despite having a focus on app reviews analysis, Dabrowski et al.~\cite{appreviewssurvey1} show that, for 2022, 59\% of works were already using ML approaches; and Genc-Nayebi and Abran \cite{appreviewssurvey2} highlight that there is a differential intention between users and developers when requesting information from reviews that should be considered in future approaches.}

Reliable LLM support in SE requires evaluation methods beyond code-centric/\rev{single-task aggregated rankings. SELU is a curated SE textual artifacts NLU benchmark that} enables a systematic evaluation of models in a variety of scenarios, revealing strengths and weaknesses that remain hidden in isolated single-task studies. \rev{As GLUE~\cite{glue}/SuperGLUE~\cite{superglue}, SELU is curated due to its diverse task selection, established data sources, and standardized evaluation framework.}


\section{Benchmark Description}
\label{s:benchmark}

In the following, we define the scope and structure of SELU. We begin by motivating our task selection criteria. Then, we describe the pre-processing pipeline and splitting strategy with specific considerations for some tasks. Together, these subsections establish the \rev{22 SE textual artifacts NLU tasks that are included in SELU}, the main characteristics of their associated datasets, and the preparation steps that underpin our evaluation framework.

\subsection{Task Selection}
\label{ss:datasets}

\begin{figure}[ht]
\centering
\caption{Distribution of published works per SE life cycle activity. Adapted from Hou et al.~\cite{llm4se2}.}
\label{fig:focus_research}
\includegraphics[width=0.4\textwidth]{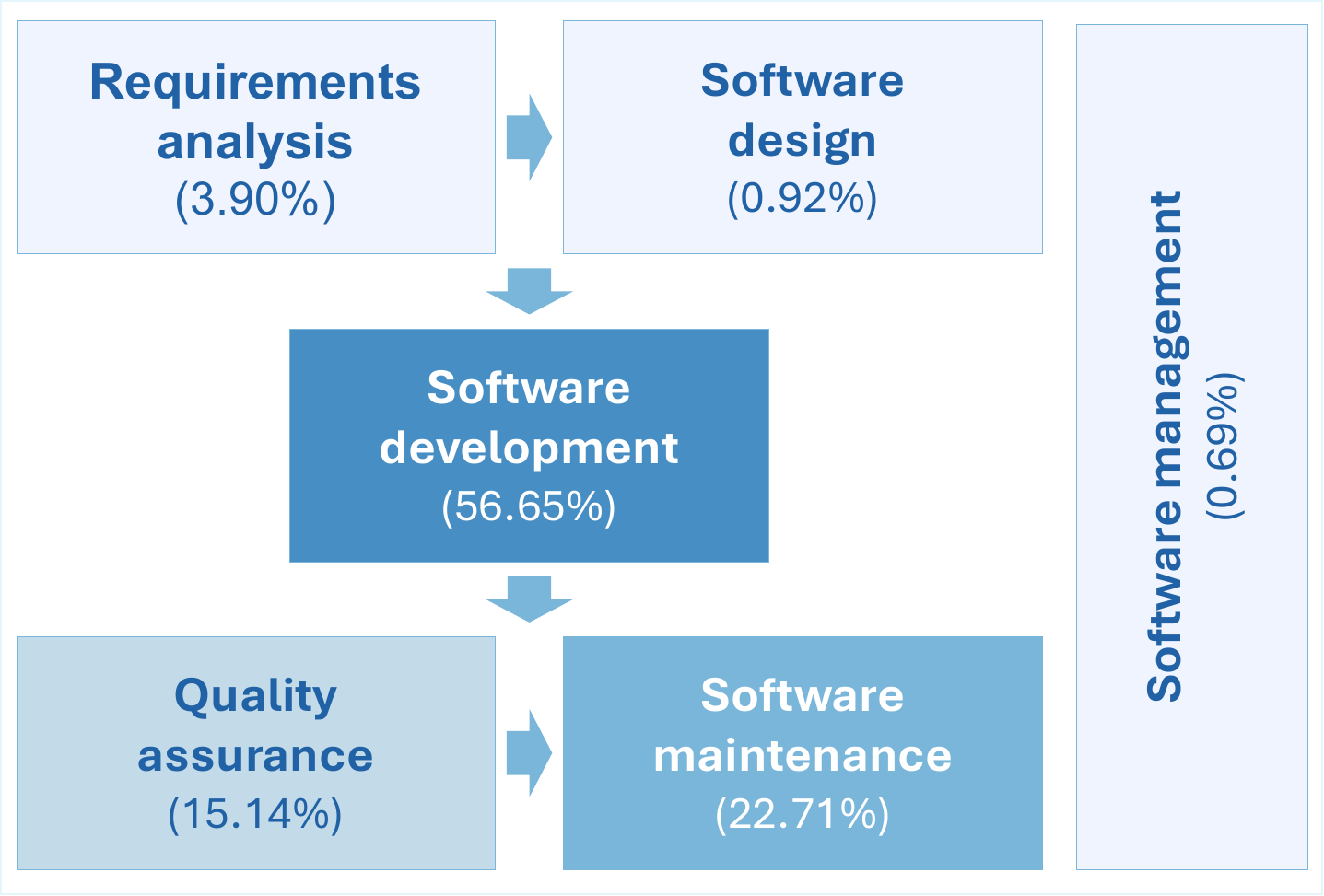}
\end{figure}

To select the tasks to include in SELU, we mainly consider the literature review by Hou et al.~\cite{llm4se2}. In this, after collecting 395 works from January 2017 to January 2024, the authors analyze each one from different dimensions such as publication year, venue, LLMs, data preparation, training, and prompting techniques implemented\rev{. Each work is also framed in the context of a task relevant for an activity throughout the SE life cycle. This distribution can be appreciated in Figure~\ref{fig:focus_research}}, where about 79\% of the works address the application of LLMs only in software development and maintenance-related tasks.

\begin{figure}[ht]
\centering
\caption{Distribution of tasks per SE life cycle activity in SELU.}
\label{fig:tasks_per_activity}
\includegraphics[width=0.4\textwidth]{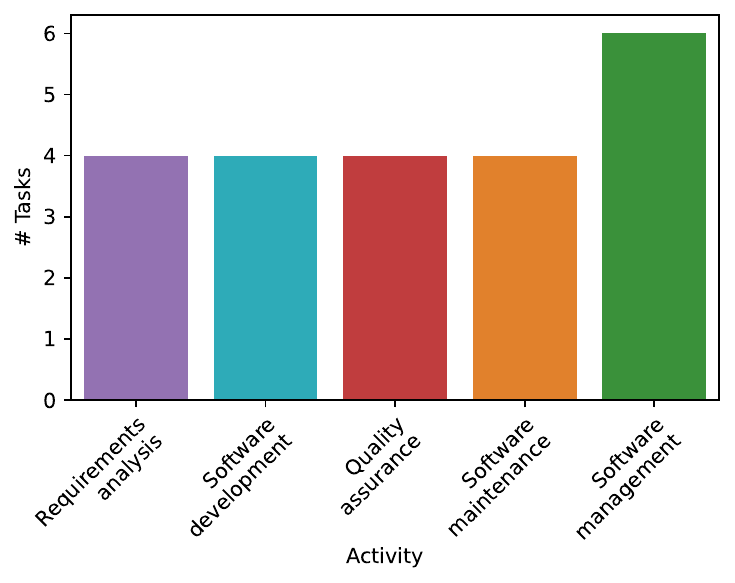}
\end{figure}

Due to the scope of our work, we discard \rev{from Hou et al. \cite{llm4se2} and other selected studies} any kind of code-\rev{centric} and generation task. In other words, we focus only on \rev{SE textual artifacts} NLU tasks, taking advantage of their well-defined labels and established evaluation framework. After removing some duplicated tasks and those for which the dataset is not publicly available, we proceed with a total of \rev{22} tasks of type classification (including binary, multi-class and multi-label), regression, Named Entity Recognition (NER), and Masked Language Modeling (MLM). Hou et al.~\cite{llm4se2} do not assign all tasks to a SE life cycle activity. Then, we validate and complement these assignments\rev{, but also assign each task to a task family (code-adjacent metadata, core non-code, or developer communication) to distinguish the different roles of natural language. We report the assignment rationale in Appendix~\ref{a:taskinventory}}. Table~\ref{tbl:tasks} lists the tasks accompanied by a brief description, the number of instances and targets each task has, \rev{and a colored shape representing the SE life cycle activity and task family to which each task is assigned}. Complementarily, Figure~\ref{fig:tasks_per_activity} shows the distribution of tasks per SE life cycle activity and \rev{Table~\ref{tbl:tasks_per_family} details, in addition to the task family, the data source from which the dataset for each task is collected.}

\newcommand{\TaskSquare}[2]{%
  #2\ \tikz[baseline=0ex]{\fill[#1] (0,0) rectangle (1.1ex,1.1ex);}%
}

\newcommand{\TaskCircle}[2]{%
  #2\ \tikz[baseline=0ex]{%
    \fill[#1] (0.55ex,0.55ex) circle[radius=0.55ex];%
  }%
}

\newcommand{\TaskTriangle}[2]{%
  #2\ \tikz[baseline=0ex]{%
    \fill[#1] (0,0) -- (1.1ex,0) -- (0.55ex,1.1ex) -- cycle;%
  }%
}

\begin{table*}[ht]
\centering
\caption{Tasks included in SELU. \rev{Colors and shapes represent the SE life cycle activity and task family, respectively}.}
\label{tbl:tasks}
\begin{tabular}{cllrcc}
\toprule
                                          & \textbf{Task ID}                                           & \textbf{Task definition}                                         & \textbf{Instances} & \textbf{Targets} & \textbf{Ref.}                                           \\
\midrule
\parbox[t]{2mm}{\multirow{7}{*}{\rotatebox[origin=c]{90}{Binary}}} 
      & \TaskTriangle{tab10red}{\textit{bug\_issue}}                   & Is the issue reporting a bug?                                 & 38,219             & 2 & \cite{bug_issue}                                          \\
                                          & \TaskSquare{tab10purple}{\rev{\textit{functional\_requirement}}}            & \rev{Does the requirement include some functional aspect?}             & \rev{956}                & 2 & \cite{functional_requirement}                                   \\
                                          & \TaskCircle{tab10green}{\textit{incivility}}                   & Does the text show unnecessary rude behavior?             & 1,546              & 2 & \cite{incivility+tone_bearing1,incivility+tone_bearing2}  \\
                                          & \TaskSquare{tab10purple}{\rev{\textit{quality\_requirement}}}            & \rev{Does the requirement include some quality aspect?}             & \rev{956}                & \rev{2} & \cite{functional_requirement}                                   \\
                                          & \TaskTriangle{tab10red}{\rev{\textit{safety\_issue}}}            & \rev{Is the issue reporting safety-related concerns?}             & \rev{1,916}                & \rev{2} & \rev{\cite{safety_issue}}                                   \\
                                          & \TaskSquare{tab10purple}{\rev{\textit{security\_requirement}}}            & \rev{Does the requirement include some security aspect?}             & \rev{510}                & \rev{2} & \rev{\cite{security_requirement}}                                   \\
                                          & \TaskCircle{tab10green}{\textit{tone\_bearing}}                & Does the text have an unnecessarily disrespectful tone?            & 6,597              & 2 & \cite{incivility+tone_bearing1,incivility+tone_bearing2}  \\
\midrule
\parbox[t]{2mm}{\multirow{7}{*}{\rotatebox[origin=c]{90}{Multi-class}}} & \TaskCircle{tab10green}{\textit{closed\_question}}             & Which is the reason for closing the question after moderation?                                 & 140,272            & 5 & \cite{closed_question}                                    \\
                                          & \TaskTriangle{tab10orange}{\textit{commit\_intention}}               & Is the commit perfecting or correcting the code?             & 2,533              & 3 & \cite{commit_intention}                                      \\
                                          & \TaskTriangle{tab10orange}{\rev{\textit{issue\_intention}}}                  & \rev{Which is the intention expressed in the issue?}            & \rev{6,375}            & \rev{7} & \rev{\cite{issue_intention}}                                         \\
                                          & \TaskTriangle{tab10orange}{\textit{issue\_type}}                  & Is the issue related to a bug, an enhancement or a question?            & 803,417            & 3 & \cite{issue_type}                                         \\
                                          & \TaskCircle{tab10green}{\textit{question\_quality}}            & Is the question of good quality or does it require moderation?                             & 60,000             & 3 & \cite{question_quality}                                   \\
                                          & \TaskSquare{tab10orange}{\rev{\textit{review\_type}}}                    & \rev{Which is the intention expressed in the app review?}                & \rev{1,390}             & \rev{4} & \rev{\cite{review_type}}                                          \\
                                          & \TaskCircle{tab10green}{\textit{sentiment}}                    & Which is the sentiment expressed in the text?                & 13,144             & 3 & \cite{sentiment}                                          \\
\midrule
\parbox[t]{2mm}{\multirow{5}{*}{\rotatebox[origin=c]{90}{Multi-label}}} & \TaskTriangle{tab10blue}{\textit{comment\_type\_java}}          & Which kind of contents are detailed in the code comment? (Java)            & 9,339              & 7 & \cite{comment_type}                                       \\
                                          & \TaskTriangle{tab10blue}{\textit{comment\_type\_pharo}}         & Which kind of contents are detailed in the code comment? (Pharo)            & 1,587              & 7 & \cite{comment_type}                                       \\
                                          & \TaskTriangle{tab10blue}{\textit{comment\_type\_python}}        & Which kind of contents are detailed in the code comment? (Python)            & 2,290              & 5 & \cite{comment_type}                                       \\
                                          & \TaskSquare{tab10red}{\textit{review\_aspect}}               & Which aspects are involved in the API review?                  & 4,522              & 11 & \cite{review_aspect}                                     \\
                                          & \TaskSquare{tab10red}{\textit{smell\_doc}}                   & Which does the API documentation smell like?                            & 1,000              & 5 & \cite{smell_doc}                                          \\
\midrule
Regr.                                     & \TaskSquare{tab10green}{\textit{story\_points}}                & What is the effort estimated for the development task?                   & 23,313             & [1-96] & \cite{story_points}                                  \\
\midrule
NER                                       & \TaskSquare{tab10blue}{\textit{se\_entities}}                 & What kind of SE terminology is discussed in the text?            & 2,718              & 20 & \cite{se_entities}                                       \\
\midrule
MLM                                       & \TaskSquare{tab10purple}{\textit{requirement\_completion}}      & How to fill the specification with the proper user action?           & 40*                & * & \cite{requirement_completion} \\
\bottomrule
\end{tabular}
\end{table*}

\begin{table*}[ht]
\centering
\caption{\rev{Data sources and task family for the tasks included in SELU.}}
\label{tbl:tasks_per_family}
\begin{tabular}{p{5.5cm}lp{8cm}}
\toprule
\textbf{Data source} & \textbf{Task family} & \textbf{Task IDs}\\
\midrule
API documentations & \TaskTriangle{black}{Code-adjacent metadata} & \textit{smell\_doc} \\
\midrule
App stores (e.g., Apple Store and Google Play) & \TaskSquare{black}{Core non-code} & \textit{review\_type} \\
\midrule
\multirow{2}{*}{Developer forums (e.g., Stack Overflow)} & \TaskSquare{black}{Core non-code} & \textit{review\_aspect}, \textit{se\_entities}\\
 & \TaskCircle{black}{Developer communication} & \textit{closed\_question}, \textit{question\_quality}, \textit{sentiment} \\
\midrule
\multirow{3}{*}{\parbox[c][1.6cm][c]{5.5cm}{Issue tracking systems (e.g., GitHub, Jira)}} & \TaskTriangle{black}{Code-adjacent metadata} & \textit{bug\_issue}, \textit{commit\_intention}, \textit{comment\_type\_java}, \textit{comment\_type\_pharo}, \textit{comment\_type\_python},  \textit{issue\_intention}, \textit{issue\_type}, \textit{safety\_issue} \\
 & \TaskSquare{black}{Core non-code} & \textit{story\_points}\\
 & \TaskCircle{black}{Developer communication} & \textit{incivility}, \textit{sentiment}, \textit{tone\_bearing}\\
\midrule
Mailing lists (e.g., Linux Kernel) & \TaskCircle{black}{Developer communication} & \textit{incivility}, \textit{tone\_bearing} \\
\midrule
\multirow{1}{*}{\parbox[c][0.6cm][c]{5.5cm}{Requirements specifications}} & \TaskSquare{black}{Core non-code} & \textit{functional\_requirement}, \textit{quality\_requirement}, \textit{requirement\_completion}, \textit{security\_requirement} \\
\bottomrule
\end{tabular}
\end{table*}

\rev{SELU task coverage is uneven in terms of task type and data sources. This means that classification tasks are highly predominant in comparison with any other task type, and most of their datasets come from issue tracking systems and developer forums, which is consistent with the fact that it is where most  public data is found.} We do not find any NLU tasks related to software design \rev{with a publicly available dataset. While representative of the findings from literature reviews, these become valuable opportunities to address as future work that requires significant data collection and labeling efforts.}

\rev{Some excluded but still relevant tasks include: (i) Prioritization of accessibility-related user reviews~\cite{reviews_prioritization}), (ii) ambiguity detection in requirements~\cite{ambiguity_requirement}), (iii) requirements risk prediction~\cite{risk_requirement}), and (iv) traceability link recovery between software architecture documentation and models~\cite{traceability_recovery}). The main reason for rejecting them is that their publicly available datasets have very few instances, while the latter two require in addition a label harmonization treatment necessary for NLU.}

\subsection{Data Preparation}
\label{ss:dataprep}

We implement different pre-processing pipelines tailored to each task with some common steps including: (1) Normalizing white spaces, (2) removing any kind of markdown and HTML styling, and (3) masking by some common regex patterns, such as URLs, hashes, user mentions, and code blocks. The rationale behind masking code blocks is that we decide to evaluate LLMs on their capabilities to solve \rev{SE textual artifacts-related} tasks, based solely on the context that natural text can provide. However, we also understand that context provided in the form of source code can be critical to success in some tasks, so comparing these two scenarios represents an opportunity to extend this work.

For the NER (\textit{se\_entities}) task, we do not apply any specific pre-processing pipeline because Tabassum et al. \cite{se_entities} already provide the dataset cleaned and tokenized at word level with their corresponding tags. And for the MLM (\textit{requirement\_completion}) task, because Luitel et al. \cite{requirement_completion} only provide the dataset in raw format (that is, 40 requirements specifications without any kind of pre-processing or masking), we define a simpler version of this task where the goal is to predict potential user actions in the specification. We use Part-Of-Speech (POS) verbs as a proxy for user actions, leveraging spaCy \cite{spacy} for word-level tokenization and POS tagging. Because the raw dataset is mostly plain text, we do not consider applying additional pre-processing.


Finally, to facilitate the experimentation process, we consistently apply a 80\%/20\% hold-out splitting strategy \rev{with a fixed global seed for all datasets. For binary and multi-class classification tasks, we use standard stratification. For multi-label, we use iterative stratification to preserve a similar representativeness and co-occurrence of labels in the training and testing subsets. For the other tasks, we do not apply stratification, but manually check the resulting subsets for severe drifts. As part of the prompting protocol, we draw 3 shots per testing instance from the training subset, stratified by class/label (when feasible), to be used across all experiments. The training and testing subsets as well as the shots for all tasks are included in our replication package~\cite{replicationpackage}.}

\section{Methodology}
\label{s:methodology}

In this section, we present the experimental framework used to evaluate whether a set of open-source LLMs and proprietary prompted alternatives perform consistently across the \rev{22 SE textual artifacts NLU tasks included in SELU}. First, we list the selected LLMs and describe how they are fine-tuned and prompted to handle different task types. We then detail the construction of baseline models, evaluation metrics, and \rev{the Bayesian analysis for pairwise model comparison across tasks}. Finally, we outline the environment that supports our experiments.

\subsection{LLMs and Baselines}
\label{ss:models}

A summary of the LLMs selected to run SELU can be found in Table \ref{tbl:llms}. We start with a diversity of well-established open-source LLMs such as BERT \cite{bert}, RoBERTa~\cite{roberta}, GPT-2~\cite{gpt2}, and T5 \cite{t5}, but given the rapid progress of the field and the constant release of innovative alternatives, we also consider ModernBERT \cite{modernbert} and Llama 3.2 \cite{llama32}. Complementarily, we include domain-adapted LLMs such as CodeBERT \cite{codebert}, CodeLlama \cite{codellama}, StarCoder2 \cite{starcoder2}, and CodeT5+ \cite{codet5plus}. All of these 22 open-source LLMs cover the main architectures (encoder-only, decoder-only, and encoder-decoder) and a range from 60.5 million to 7 billion parameters. We do not experiment with LLMs larger than 7 billion parameters because of our compute budget.

\begin{table}[ht]
\centering
\caption{LLMs selected for evaluation. \rev{First compartment: Generalist open-source. Second compartment: Domain-adapted open-source. Third compartment: Generalist proprietary.}}
\label{tbl:llms}
\begin{tabular}{lcccc}
\toprule
\textbf{Model} & \textbf{\textbf{Size}} & \textbf{Architecture} &  \textbf{Source} \\
\midrule
BERT base           & 110M                   & encoder-only          & \cite{bert} \\
BERT large          & 340M                   & encoder-only          & \cite{bert} \\
RoBERTa base        & 125M                   & encoder-only          & \cite{roberta} \\
RoBERTa large       & 355M                   & encoder-only          & \cite{roberta} \\
ModernBERT base     & 150M                   & encoder-only          & \cite{modernbert} \\
ModernBERT large    & 396M                   & encoder-only          & \cite{modernbert} \\
GPT-2 small         & 117M                   & decoder-only          & \cite{gpt2} \\
GPT-2 medium        & 345M                   & decoder-only          & \cite{gpt2} \\
GPT-2 large         & 774M                   & decoder-only          & \cite{gpt2} \\
GPT-2 xl            & 1.5B                   & decoder-only          & \cite{gpt2} \\
Llama 3.2 1B        & 1B                     & decoder-only          & \cite{llama32} \\
Llama 3.2 3B        & 3B                     & decoder-only          & \cite{llama32} \\
T5 small            & 60.5M                  & encoder-decoder       & \cite{t5} \\
T5 base             & 223M                   & encoder-decoder       & \cite{t5} \\
T5 large            & 738M                   & encoder-decoder       & \cite{t5} \\
T5 3B               & 3B                     & encoder-decoder       & \cite{t5} \\
\midrule
CodeBERT base       & 125M                   & encoder-only          & \cite{codebert} \\
CodeLlama 7B        & 7B                     & decoder-only          & \cite{codellama} \\
StarCoder2 3B       & 3B                     & decoder-only          & \cite{starcoder2} \\
StarCoder2 7B       & 7B                     & decoder-only          & \cite{starcoder2} \\
CodeT5+ 220M        & 220M                   & encoder-decoder       & \cite{codet5plus} \\
CodeT5+ 7700M       & 770M                   & encoder-decoder       & \cite{codet5plus} \\
\midrule
GPT-4o              & -                      & decoder-only          & \cite{gpt4o} \\
Claude 3.7 Sonnet   & -                      & decoder-only          & \cite{claude37} \\
\bottomrule
\end{tabular}
\end{table}

It is important to note that code-focused pre-training, the main domain adaptation technique applied in the selected models, implies that LLMs have ingested tons of source code \cite{codellama,starcoder2}\rev{, although also with data from publicly available sources such as issue tracking systems and developer forums}. This means that their internal representations, tokenization schemes, and optimization objectives are heavily biased toward syntactic and semantic patterns of programming languages. As a result, gains observed in purely code understanding or generation may not directly translate to other kind of SE tasks (e.g., \textit{functional\_requirement} or \textit{story\_points}), where linguistic diversity and domain-specific terminology differ significantly from typical code corpora. By evaluating these models, we aim to quantify the extent to which this type of domain adaptation offers tangible benefits (or introduces trade-offs) in diverse SE textual artifact scenarios.


We also run SELU on two proprietary LLMs \rev{via zero and 3-shot prompting:} GPT-4o \cite{gpt4o} and Claude 3.7 Sonnet \cite{claude37}. Although the actual number of parameters of these models is unknown, they are generally considered to be at least one magnitude larger than the selected open-source LLMs. \rev{Practitioners tend to use proprietary generative alternatives like these for NLU tasks taking advantage of their capacity to follow instructions without requiring specific training. This motivates us to keep both kinds of results together in some analysis to better understand the trade-offs between task-specific adaptation and in-context learning. It is also possible to prompt open-source, decoder-only LLMs such as GPT-2, Llama 3.2, CodeLlama, and StarCoder2. However, even their instruction-tuned versions (when available) do not follow the instructions from prompts reliably, generating high rates of invalid outputs. For this reason, we do not include the results for these experiments, but they can be found in our replication package~\cite{replicationpackage} for selected tasks.}

Finally, we train two baseline models on each SELU task using classical ML approaches such as TF-IDF+XGBoost \cite{xgboost} and FastText \cite{fasttext}. By including them, we measure the performance improvement that LLMs can offer to compensate for their increasing complexity.

\subsection{Task Configuration}
\label{ss:taskconfig}

\begin{figure*}[ht]
\centering
\caption{LLM heads for different task types.}
\label{fig:heads}
\includegraphics[width=0.93\textwidth]{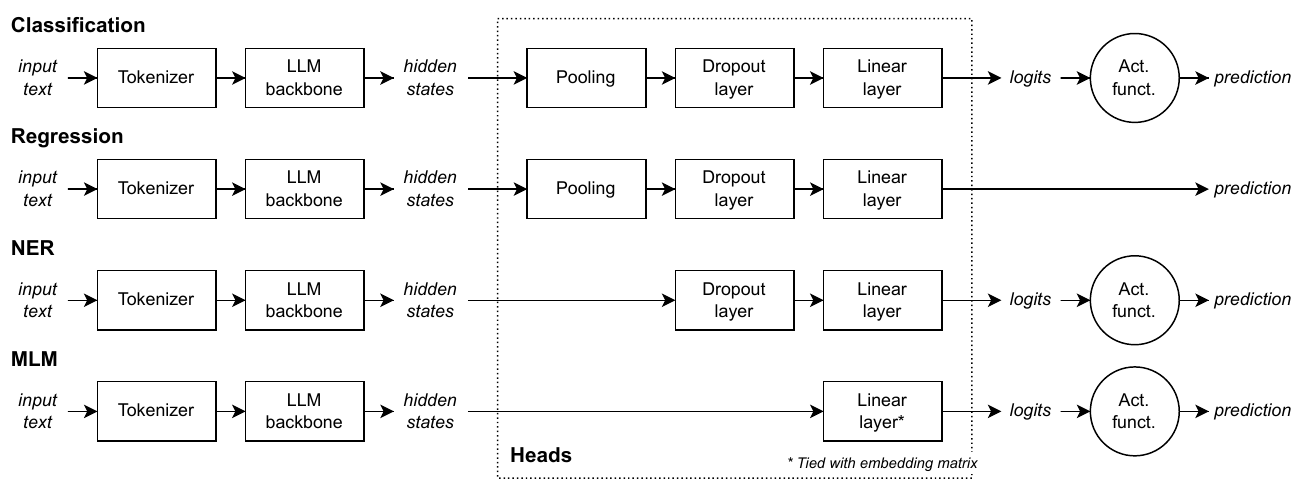}
\end{figure*}

LLMs are not configured by default to solve NLU tasks such as classification, regression, and NER. For this purpose, we extend them with additional trainable components (a.k.a. heads) tailored to each task type. This extension is shown at a high level of abstraction in Figure \ref{fig:heads}, where the pooling layers are intended to extract relevant hidden states (e.g., from the [CLS] token in encoder-only LLMs), the dropout layers control overfitting particularly when fine-tuning on small datasets, and the linear layers map hidden states to a task-specific output space. Finally, at inference time, the activation functions transform logits into predictions in the target range.

As mentioned above, for NER and MLM tasks, we need to handle targets at token level. Independently of any previous tokenization process (e.g., the word-level tokenization applied to the \textit{se\_entities} dataset by Tabassum et al. \cite{se_entities}), each LLM applies additional sub-word tokenization before feeding the input into the network. To avoid breaking the relationship between tokens and targets, we apply an alignment process to keep both data structures the same size.

In particular for our custom definition of the \textit{requirement\_completion} task and according to what is already introduced in Subsection~\ref{ss:dataprep}, this is not intended to evaluate LLMs on their capacity to predict any kind of token, but POS verbs. To meet this need, we implement a custom data collector that applies token masking only on this type of token with a probability of 50\%. The reason for having a higher masking probability compared to a standard MLM task \cite{bert} is that POS verbs only represent about 9\% of the tokens.

Other considerations for NER and MLM tasks are on the side of the architectures that support them. NER can be formulated either as a token-level classification or as a text‐to‐text generation problem. Decoder‐only LLMs lack a mechanism to produce token-level labels without post‐processing of generated text, making them ill‐suited for this task \cite{gpt-ner}. Apart from that, MLM requires the ability to predict individual masked tokens based on a full bidirectional context. Decoder‐only and standard encoder–decoder architectures lack an isolated mechanism for masked token prediction, and decoder‐only models also cannot access bidirectional context.

In summary, while the classification and regression tasks are used to evaluate the entire set of LLMs, the NER task is intended for the encoder-only and encoder-decoder subset, while the MLM task is only supported for the encoder-only LLMs. Finally, to the best of our knowledge, there is no standard definition of NER and MLM using classical ML approaches; therefore, baseline models are also not considered for these tasks.

\subsection{Training, Prompting and Evaluation}
\label{ss:training}

\rev{We determine a common set of hyper-parameters from initial experimentation, looking for a consistent convergence behavior across all LLMs. Moreover, we do not perform additional hyper-parameter tuning to ensure fairer comparisons.} We use a batch size of $64$, and reduce it to $32$ or $16$ in combination with gradient accumulation steps of $2$ or $4$ to avoid memory errors, particularly with billion parameter LLMs. We run the fine-tuning experiments for \rev{$10$ epochs (enough for most of them to reach convergence)}, setting an evaluation interval to $10\%$ of the total calculated steps with an early stopping criterion of $3$ intervals. For the NER and MLM tasks, we use $50$ epochs due to the small size of their datasets\rev{, especially in relation to the number of targets}.

\rev{Since LLMs trained on small datasets require less iterations (model updates) for the same number of epochs, we use a learning rate of $1\mathrm{e}{-4}$ for tasks with $<10K$ instances; otherwise, $1\mathrm{e}{-5}$. We find this policy appropriate as it guarantees better convergence for most of the tasks within our study; however, for Llama 3.2 and StarCoder2 series, our evidence suggests that it is better to use a learning rate of $1\mathrm{e}{-5}$ in all cases because they exhibit unstable convergence under larger learning rates. The learning rate is modified by a warm-up period of $10\%$, then decreasing linearly to $0$. Other key hyper-parameters include} AdamW as optimizer, weight decay of $0.01$, and BFloat16 mixed precision, to optimize our compute budget. 

For prompting, we design \rev{task-tailored templates that adapt to zero-shot and 3-shot strategies following} a common base prompt template, as shown in Figure \ref{fig:prompt}. Because studying the effectiveness of state-of-the-art prompting techniques is out-of-scope in this work, we design the base prompt template to make it standard and straightforward to define\rev{, requiring aspects such as context, task definition, output format, class/label schema, and shots. To process responses, we apply regex validation, treat invalid outputs as errors and log rates. All the prompt templates are available in our replication package~\cite{replicationpackage}.} 

\begin{figure*}[ht]
\centering
\caption{\rev{Base prompt template generally applied across tasks. New lines partially omitted to reduce space.}}
\label{fig:prompt}
\begin{tcolorbox}[
    colback=gray!15,
    colframe=gray!50,
    enhanced,
    rounded corners,
    boxrule=0.5pt,
    title={Base prompt template},
    width=0.9\textwidth,
    boxsep=3pt,    
    left=3pt,      
    right=3pt,
    top=0pt,
    bottom=0pt,
    before skip=2pt, 
    after skip=3pt,   
    fontupper={\fontsize{8pt}{10pt}\selectfont\ttfamily},
    fonttitle={\fontsize{8pt}{10pt}\selectfont\ttfamily}
]
\begin{verbatim}
In the context of <context>, <task>. <answer format>.
The meaning of each <class/label> is as follows: <CLASS/LABEL DEFINITIONS>
- - -
Below are some examples to guide you in the <classification/estimation> task:
- EXAMPLE N:
<ENTITY_TYPE>: {shot_N}
<CLASS/LABELS>: {target_N}
- - -
<ENTITY_TYPE> TO <CLASSIFY/ESTIMATE>: {text}
\end{verbatim}
\end{tcolorbox}












\end{figure*}

For training the TF-IDF+XGBoost~\cite{xgboost} baseline models, we select the best candidate by experimenting with two key hyper-parameters: maximum depth, with values ranging from $3$ to $10$, and minimum child weight, ranging from $1$ to $7$. Similarly, for FastText~\cite{fasttext}, we experiment with learning rate and number of epochs, ranging from $1\mathrm{e}{-6}$ to $1\mathrm{e}{-1}$ with exponential steps, and from $5$ to $20$, respectively. Nevertheless, these ranges cover shallow and deep trees exhibiting different degrees of robustness to overfitting at the individual-tree level within the XGBoost ensemble, as well as the commonly used range of learning rates for FastText. These settings are therefore expected to yield strong results, and we do not observe systematic issues during training, as most configurations converge satisfactorily and yield well-defined confusion matrices in a reasonable time. Because FastText does not support regression targets, we only use it to build vector representations of tokens that are then averaged and used as features to train a linear regressor. Due to its simplicity, we do not perform model selection (i.e., conduct experiments with different hyper-parameters) for the FastText approach on this task. 

\rev{Although we compute and publish different evaluation metrics tailored to each task type in our replication package~\cite{replicationpackage}, for the purpose of applying the Bayesian analysis detailed in Subsection~\ref{ss:statanalysis} and other comparisons,} we select F1-macro \cite{f1macro} as the \rev{reporting} metric for classification tasks due to its robustness in terms of the number of classes/labels in binary, multi-class, and multi-label settings. It computes the F1-score independently for each class based on its correct and incorrect predictions and then averages them equally, providing a balanced view of model performance regardless of potential class-level imbalance in some datasets. The formula is given in Equation (\ref{eq:f1macro}), where $C$ is the number of classes, and $F1_i$, $TP_i$, $FP_i$, and $FN_i$ are the F1-score, the number of true positives, false positives, and false negatives for class $i$, respectively.

\begin{equation}
\label{eq:f1macro}
\begin{aligned}
\text{F1-macro} 
&= \frac{1}{C} \sum_{i=1}^{C} \frac{2 \, TP_i}
{2 \, TP_i + FP_i + FN_i}
\end{aligned}
\end{equation}

For regression, we use Symmetric Mean Absolute Percentage Error (SMAPE) \cite{smape}. It expresses the error as a percentage, adjusted to be symmetric by dividing the absolute difference by the average of actual and predicted values, making it more robust to scale and zero values than traditional MAPE. The formula is given in Equation (\ref{eq:smape}), where $N$ is the number of instances, and $\hat{y}_j$ and $y_j$ are the predicted and actual values for instance $j$, respectively. Similarly to F1-macro, SMAPE ranges from $0$ to $1$ with lower values indicating better performance. To align performance scores for subsequent statistical analysis, we invert SMAPE by computing $1 - \text{SMAPE}$, making scores closer to $1$ always represent better performance.

\begin{equation}
\label{eq:smape}
\begin{aligned}
\text{SMAPE} 
&= \frac{1}{N} \sum_{j=1}^{N} \frac{|\hat{y}_j - y_j|}{\frac{1}{2} \left( |\hat{y}_j| + |y_j| \right)}
\end{aligned}
\end{equation}

The NER task can be addressed as a special case of classification, but instead of using F1-macro as the \rev{reporting} evaluation metric, we use F1-micro, which measures overall performance by aggregating true positives, false positives, and false negatives across all tokens and classes. In token classification tasks, it evaluates how well the model performs at token level by treating each token equally, regardless of its class. The formula is given in Equation (\ref{eq:f1micro}), where $C$ is the number of classes, and $TP_i$, $FP_i$, and $FN_i$ are the number of true positives, false positives, and false negatives for class $i$, respectively.

\begin{equation}
\label{eq:f1micro}
\begin{aligned}
\text{F1-micro}
&= \frac{2 \, \displaystyle\sum_{i=1}^{C} TP_i}
{2 \, \displaystyle\sum_{i=1}^{C} TP_i + \displaystyle\sum_{i=1}^{C} FP_i + \displaystyle\sum_{i=1}^{C} FN_i}
\end{aligned}
\end{equation}

Finally, for the MLM task, we select the standard accuracy, which measures the proportion of correctly predicted masked tokens over the total number of masked ones, providing a straightforward evaluation of how often the model is right. The formula is given in Equation (\ref{eq:accuracy}), where $N$ is the number of masked tokens, and $\hat{y}_t$ and $y_t$ are the predicted and actual tokens at position $t$, respectively. $\indicator[\cdot]$ is the indicator function, which equals $1$ if the prediction is correct and $0$ otherwise.

\begin{equation}
\label{eq:accuracy}
\begin{aligned}
\text{Accuracy} = \frac{1}{N} \sum_{t=1}^{N} \indicator[\hat{y}_t = y_t]
\end{aligned}
\end{equation}

We are aware that these metrics, despite being in the same range of values, represent different aspects of error and have a variety of implications depending on the task type \rev{that should be considered when analyzing each task independently}. For the purpose of model comparison, we treat values derived from these calculations equitably as `performance score' of a given model on a certain task.

\subsection{Bayesian analysis}
\label{ss:statanalysis}

To evaluate whether models, including both LLMs and baselines, perform consistently across classification and regression tasks included in SELU, following the recommendations of Benavoli et al. \cite{bayesian-tutorial}, we first use the Shapiro–Wilk test \cite{shapiro-wilk} to check whether performance scores per model are normal. If this is the case, we then compute the means, standard deviations, and confidence intervals (with Bonferroni correction \cite{bonferroni-correction} to maintain a 95\% family-wise confidence level) and rank the models by mean performance. For this analysis, and according to what is already introduced in Subsection \ref{ss:taskconfig}, we do not include the NER and MLM tasks because they are not supported by \rev{decoder-only LLMs and our baselines}.

Subsequently, we define a Region of Practical Equivalence (ROPE) as $\pm 0.1 \cdot d$, where $d$ is the Cohen's effect size \cite{cohens-effect} defined in Equation (\ref{eq:effect_size}). For any pair of models $A$ and $B$, $\mathit{Perf}_{A}$ and $\mathit{Perf}_{B}$ are vectors with the respective performance scores, $s$ is the pooled standard deviation, and \rev{$n_{A} = n_{B} = 20$} correspond to the number of tasks. We omit the treatment of non-normal data for brevity, since, as shown in Section \ref{s:results}, all populations pass the normality check.

\begin{gather}
\label{eq:effect_size}
d = \frac{\mathrm{mean}(\mathit{Perf}_{A}) - \mathrm{mean}(\mathit{Perf}_{B})}{s} \\[1ex]
s = \sqrt{\frac{(n_A-1)\,\mathrm{std}(\mathit{Perf}_{A})^2 + (n_B-1)\,\mathrm{std}(\mathit{Perf}_{B})^2}{n_A + n_B - 2}}
\notag
\end{gather}

As a final step, we apply the Bayesian signed-rank test~\cite{bayesian-test} for pairwise model comparison. This test returns three posterior probabilities, $\mathbb{P}(A > B)$, $\mathbb{P}(A = B)$, and $\mathbb{P}(B > A)$ for `A outperforming B,' `A and B being practically equivalent,' and `B outperforming A,' respectively. We conclude that $A$ outperforms $B$ if $\mathbb{P}(A > B) \ge 0.95$, that they are equivalent if $\mathbb{P}(A = B) \ge 0.95$, and that $B$ outperforms $A$ if $\mathbb{P}(B > A) \ge 0.95$; otherwise, the result is inconclusive. 

\subsection{Hardware and Libraries}

All our experiments are run on a server with 8 NVIDIA A100 GPUs. To enable full fine-tuning of LLMs with up to 7 billion parameter, we set up DeepSpeed (stage 2) \cite{deepspeed}, which shards optimizer states and gradients across parallel workers (i.e., GPUs), in combination with Flash attention 2 \cite{flashattn2} when the model supports it. On top of these, we use the LLM implementations from Transformers \cite{transformers}; and Scikit-Learn \cite{sklearn} and Seqeval \cite{seqeval} to compute evaluation metrics. As already mentioned, we use XGBoost \cite{xgboost} and FastText \cite{fasttext} to train the baselines, and the Open AI and Anthropic APIs to query proprietary LLMs. Finally, we use Autorank \cite{autorank} for Bayesian analysis.

\section{Results}
\label{s:results}

In the following, we report the empirical outcomes of evaluating different models on SELU. After fine-tuning the open-source LLMs, prompting the proprietary alternatives and training the baselines, we rank them by mean performance score, apply the Bayesian signed‐rank test, and establish distinguishable model tiers. Then, we analyze how key performance drivers\rev{, task family and task type} affect results both independently and considering some interactions. Detailed results per model and task are provided in Appendix~\ref{a:results-details}.

\subsection{Ranking by Mean Performance and Model Tiers}

We first check the normality of the performance scores per model. As a result, we fail to reject the null hypothesis that the distribution is normal for all models, allowing us to rank them by mean performance. As shown in Table \ref{tbl:model_ranking}, \rev{the LLMs from the GPT-2 and Llama 3.2 series have the highest mean performance, although we observe a large group of eighteen models with a mean performance between $0.709$ and $0.755$ that also includes StarCoder2 3B, the T5 series (except the smallest version), CodeT5+, ModernBERT, CodeLlama, and even the RoBERTa base and BERT base. After this group, there is a small gap before StarCoder2 7B (M=$0.679$) and our best performing baseline TF-IDF+XGBoost (M=$0.645$). The proprietary LLMs are in the bottom-third of the ranking, with a 3-shot performance of $0.601$ and $0.592$ for GPT-4o and Claude 3.7 Sonnet, respectively. The zero-shot performance for both LLMs is slightly worse with $0.573$ and $0.572$, respectively.}



\begin{table}[ht]
\centering
\caption{\rev{Ranking of models and Cohen's effect size ($d$) compared to the top-ranked model.}}
\label{tbl:model_ranking}
\begingroup
\setlength{\tabcolsep}{4pt} 
\begin{tabular}{p{2.6cm}ccccc}
\toprule
\textbf{Model} & \textbf{M} & \textbf{SD} & \textbf{CI} & $d$ & \textbf{Mag.} \\
\midrule
GPT-2 xl & 0.755 & 0.130 & {[}0.606, 0.903] & 0.000 & neg. \\
Llama 3.2 3B & 0.747 & 0.128 & {[}0.600, 0.893] & 0.061 & neg. \\
GPT-2 large & 0.746 & 0.141 & {[}0.586, 0.907] & 0.061 & neg. \\
GPT-2 medium & 0.740 & 0.141 & {[}0.579, 0.900] & 0.110 & neg. \\
Llama 3.2 1B & 0.739 & 0.128 & {[}0.594, 0.885] & 0.120 & neg. \\
GPT-2 small & 0.734 & 0.134 & {[}0.581, 0.887] & 0.157 & neg. \\
StarCoder2 3B & 0.734 & 0.156 & {[}0.556, 0.911] & 0.145 & neg. \\
T5 large & 0.733 & 0.154 & {[}0.557, 0.908] & 0.154 & neg. \\
CodeT5+ 770M & 0.731 & 0.139 & {[}0.573, 0.889] & 0.175 & neg. \\
ModernBERT large & 0.724 & 0.161 & {[}0.540, 0.908] & 0.208 & sml. \\
T5 base & 0.716 & 0.159 & {[}0.535, 0.898] & 0.262 & sml. \\
BERT base & 0.716 & 0.157 & {[}0.537, 0.895] & 0.267 & sml. \\
T5 3B & 0.715 & 0.162 & {[}0.530, 0.900] & 0.268 & sml. \\
CodeLlama 7B & 0.714 & 0.154 & {[}0.539, 0.889] & 0.285 & sml. \\
CodeT5+ 220M & 0.714 & 0.145 & {[}0.549, 0.879] & 0.295 & sml. \\
CodeBERT base & 0.714 & 0.156 & {[}0.536, 0.892] & 0.284 & sml. \\
RoBERTa base & 0.711 & 0.169 & {[}0.518, 0.903] & 0.291 & sml. \\
ModernBERT base & 0.709 & 0.164 & {[}0.523, 0.896] & 0.306 & sml. \\
StarCoder2 7B & 0.679 & 0.216 & {[}0.432, 0.926] & 0.422 & sml. \\
TF-IDF+XGBoost & 0.645 & 0.132 & {[}0.495, 0.795] & 0.839 & lar. \\
BERT large & 0.626 & 0.247 & {[}0.345, 0.908] & 0.649 & med. \\
T5 small & 0.625 & 0.227 & {[}0.366, 0.885] & 0.697 & med. \\
GPT-4o (3-shot) & 0.601 & 0.226 & {[}0.344, 0.859] & 0.831 & lar. \\
Claude 3.7 (3-shot) & 0.592 & 0.216 & {[}0.346, 0.838] & 0.914 & lar. \\
FastText & 0.574 & 0.179 & {[}0.371, 0.778] & 1.154 & lar. \\
RoBERTa large & 0.574 & 0.282 & {[}0.252, 0.896] & 0.822 & lar. \\
GPT-4o (zero-shot) & 0.573 & 0.235 & {[}0.305, 0.841] & 0.955 & lar. \\
Claude 3.7 (zero-shot) & 0.572 & 0.200 & {[}0.344, 0.799] & 1.085 & lar. \\
\bottomrule
\end{tabular}
\endgroup
\end{table}

Table \ref{tbl:model_ranking} also shows the standard deviation, confidence intervals, and Cohen's effect size ($d$) compared to the top-ranked model (i.e. \rev{GPT-2 xl}), showing that models with higher mean performance tend to have relatively lower across-task variance, while conversely models with lower mean performance have higher across-task variance. \rev{The fact that the effect size for any potentially significant effect within the large group of fine-tuned LLMs is at best small, may be partially due to the high across-task variance in comparison to the small differences in mean performance}. A direct consequence of this is that a substantial fraction of the posterior mass of the differences between models falls within the ROPE. In the Bayesian signed‐rank test, this translates into only moderate Bayes factors in favor of the outperforming model, rather than `decisive' evidence that can be easily observed in lower‐variance models. In other words, high across-task variance is a key driver of uncertainty. 

The posterior and decision matrices are reported in Figure~\ref{fig:posterior_matrices}, Appendix~\ref{a:results-details}. Through the analysis of the matrices, we determine four tiers of models:

\rev{\textbf{Tier 1:} GPT-2 and Llama 3.2. The Bayesian signed-rank test confirms that these models are the best performing and that is it very unlikely that other models are better. Due to the high inter-task variance and moderate sample size of 22 tasks, firm statistical claims are not possible. However, the posterior probability that GPT-2 xl outperforms the LLMs in the second tier is above 90\% for most of them. Still, for T5 large, T5 base, CodeLlama 7B, and ModernBERT large the posterior probabilities that GPT-2 xl outperforms them is only between 53\% and 71\%.}

\rev{\textbf{Tier 2:} Most other fine-tuned LLMs. The Bayesian signed-rank test confirms that the gap between these LLMs with a mean performance $\geq 0.709$ and those $\leq 0.625$ is significant.}

\rev{\textbf{Tier 3:} StarCoder2 7B and TF-IDF+XGBoost. The Bayesian signed-rank test confirms that these are better than the models ranked below and that there is at least a 69\% probability that they outperform the proprietary LLMs, with many values being above the significance threshold of 95\%.}

\rev{\textbf{Tier 4:} 3-shot and zero-shot prompting strategies; models with insufficient fine-tuning possibly requiring more careful attention to hyper-parameters, i.e., BERT large, RoBERTa large; and neural networks models that might be too small, i.e., T5 small and FastText. The Bayesian signed-rank test does not find big differences between the lowest ranked models, which is in line with the relatively small absolute difference in mean performance and high inter-task variance.}

\begin{mdframed}
\rev{Fine-tuned LLMs perform best on the SELU tasks, but there is no strong results favoring specific model sizes or architectures. Proprietary LLMs using zero-shot or 3-shot prompting strategies are inferior and even outperformed by TF-IDF+XGBoost models.}
\end{mdframed}



\subsection{Impact of Performance Drivers}

We look for additional patterns in terms of performance drivers. \rev{Overall, model size and architecture do not have a major impact on performance: while larger LLMs within the same series tend to give better results, as is shown at the top of the ranking, this is not consistently the case and even smaller LLMs such as BERT base can compete with models with several billion parameters. Whether LLMs are generalist or domain-adapted makes a difference when aggregating results by task family and task type is shown in Table~\ref{tbl:task-family-results}. For code-adjacent and core non-code tasks, we see small improvements on mean performance and variance in favor of domain adoption via code-focused pre-training.\footnote{\rev{The small mean difference for core non-code tasks may be attributed to the low performance of largest versions of BERT and RoBERTa for individual multi-label tasks (see Table~\ref{tbl:results1}). Without these outliers, no difference remains.}} However, for developer communication tasks, a similar improvement is observed this time in favor of generalist LLMs.}

\rev{While the performance across task families and LLM types (i.e., generalist and domain-adapted) is not equal, it is remarkably stable with mean performance scores between $0.670$ and $0.739$ and the mean difference on code-adjacent metadata tasks being the lowest. Furthermore, the task type (seen as a proxy of complexity) offers complementary insights. Binary classification tasks have the lowest complexity and achieve the highest mean performance; while multi-label tasks, which are the most complex variant of classification, have clearly the lowest mean performance but seem to benefit more from domain adaptation via code-focused pre-training. We cannot reliably say how regression or token-level tasks like NER or MLM behave, due to the low number of such tasks.}

\begin{mdframed}
\rev{There are no evident performance drivers other than the task type as a proxy of complexity. This motivates a more careful experiment design instead of using a specific LLM.}
\end{mdframed}

\begin{table}
\centering
\caption{\rev{Impact of task family and task type on fine-tuned LLMs, i.e., excluding baseline, zero-shot, and 3-shot results.}}
\label{tbl:task-family-results}
\begin{tabular}{lcccc}
\toprule
& \multicolumn{2}{c}{\textbf{Generalist}} &  \multicolumn{2}{c}{\textbf{Domain-adapted}} \\
\midrule
\textbf{Task family / Task type} & \textbf{M} & \textbf{SD} & \textbf{M} & \textbf{SD} \\
\midrule
Code-adjacent metadata & 0.723 & 0.154 & 0.739 & 0.148 \\
Core non-code & 0.670 & 0.234 & 0.694 & 0.196 \\
Developer communication & 0.722 & 0.122 & 0.694 & 0.131 \\
\midrule
Binary classification & 0.806 & 0.110 & 0.783 & 0.140 \\
Multi-class classification & 0.750 & 0.124 & 0.761 & 0.098 \\
Multi-label classification & 0.575 & 0.173 & 0.611 & 0.160 \\
\bottomrule
\end{tabular}
\end{table}

\section{Discussion}
\label{s:discussion}

In this work, we pose the research question: \rev{\textit{Do the selected LLMs perform consistently across different SE textual artifacts NLU tasks?}} \rev{The answer to this question has four aspects.}

\rev{First, while performance is notably variable across individual tasks, we do not observe a strong performance variation across task families. This indicates that while individual task types (and likely data quality and volume) are a factor, LLMs are equally well-suited for code-adjacent metadata, core non-code, and developer communication tasks.}

\rev{Second, the model choice seems to be a minor factor overall: While fine-tuning large decoder-only models from the GPT-2 and Llama 3.2 series yield the best mean performance, the difference to a large variety of other LLMs is relatively small. While the advantages of newer LLMs for generative tasks are well documented (see, e.g., \cite{llm4se1}), our evidence suggests that these do not necessarily translate to fine-tuning for NLU tasks, where outdated LLMs such as GPT-2, T5, and BERT remain competitive.}

\rev{Third, domain-adaption has a surprisingly low impact on the results. The largest, though still small, mean performance differences are observed in multi-label classification tasks in favor of domain adaptation, and developer communication tasks in favor of the generalist LLMs. However, our evidence is limited as we only consider domain adaption via code-focused pre-training. This warrants further investigation to better understand if domain adaption in SE yield practical advantages for non-code tasks when mostly SE textual data is used for domain adaption, as indicates early research~\cite{mosel2023}.}

\rev{Fourth, while using zero-shot or 3-shot strategies for proprietary LLMs may seem convenient, these are fairly consistently outperformed by LLMs fine-tuned on SE textual artifacts NLU tasks and they can hardly compete with simpler models such as those based on TF-IDF+XGBoost.}

\subsection{Implications for SE Practice}

Our findings inform several practical guidelines for integrating LLMs into non-code SE workflows:

\begin{itemize}
\item \textbf{Model selection:} \rev{Fine-tuning open-source LLMs is the best alternative for SE textual artifacts NLU tasks. Furthermore, our results suggest that LLMs around 1B/3B parameters perform best, but smaller ones such as BERT base can still be competitive  providing a good balance between consistent performance and efficiency.}
\item \rev{\textbf{LLMs in production}: Our results demonstrate that in NLU settings, whenever possible, fine-tuning open-source LLMs is a better alternative than prompting proprietary alternatives. Unless the gap in performance does not matter or additional development and maintenance effort is not feasible, proprietary LLMs should be avoided.}
\item \rev{\textbf{Fast prototyping:} Since proprietary LLMs operate out-of-the box through API calls, they are valuable tools for building fast prototypes, although they may not achieve the same performance or cost-effectiveness compared to running inferences on smaller open-source LLMs fine-tuned carefully on specific tasks.}
\item \textbf{Domain adaptation:} \rev{Code-focused pre-training does not yield performance gains on SE textual artifacts NLU tasks. Moreover, whether pre-training on natural language texts from the SE domain lead to better domain adoption in scenarios different from code understanding and generation is not sufficiently explored.}
\end{itemize}

These guidelines suggest that deployment strategies should be tailored to task definition and organizational constraints, rather than adopting a one-size-fits-all approach.

\subsection{Future Directions}

SELU should be broadened into \rev{four} key dimensions:

1) Extending the scope of tasks beyond NLU, in addition to covering evaluation aspects related to foundational knowledge in SE and software design. This means adding tasks such as (i) Q\&A of concepts, paradigms, and common practices, (ii) recommendation of design patterns, (iii) architectural evaluation, (iv) generation of UML diagrams, or (v) synthesis of API scaffolding, among others; moving from merely language understanding to generation tasks that mirror real-world specification and design workflows. To the best of our knowledge, the previous tasks require new data collection and labeling efforts. Furthermore, generation tasks introduce distinct methodological challenges (metric validity, hallucination/safety checks, human-in-the-loop adjudication) that differ from typical NLU protocols. However, overcoming these challenges is essential insofar as LLMs and agentic systems are nowadays considered for every aspect of software development, requiring that benchmarks such as SELU extend their coverage to comprehensively evaluate the very broad range of LLM capabilities.

2) \rev{Once the challenges behind negotiating NDAs have been overcome, including industrial/proprietary datasets, together with validation via live developer environments, represents a natural mechanism to reinforce the value of SELU from a more practical perspective. This enables assessing transferability, documenting domain shift and considering cost/latency trade-offs.}

3) Investigating domain adaptation via pre-training with a stronger focus on natural language texts from the SE domain such as design documents, requirements specifications, meeting transcripts, among others. By comparing models pre-trained on general-purpose corpora against those infused with new varieties of SE data, researchers can quantify how domain-specific priors influence both understanding and generation throughout the SE life cycle.

4) Systematically studying stability and reproducibility. Future experiments should vary random seeds and key hyper-parameters, ensuring that performance gains are robust and repeatable in practical SE deployments. Moreover, designing prompts is not trivial and temperature also has a significant impact on LLM responses, a reason why these elements have to be considered in generation tasks.

\section{Threats to Validity}
\label{s:threats}

In this section, we systematically examine potential threats to the validity \cite{research} of SELU and our experimental framework in four dimensions to provide context for the interpretation of results. By identifying these limitations, we aim to guide future work towards improving the robustness and generalizability of LLMs in SE.

\subsection{Construct Validity}


We built SELU by collecting existing datasets of different sizes, summing up million instances. These datasets come mostly from peer-reviewed works, and, to the best of our knowledge, the labels defined for each task are plausible. However, most of the processes the authors follow for data collection, data preparation, and labeling are beyond our control. The major exception is the \rev{\textit{requirement\_completion} task that we redefine as predicting} POS verbs as a proxy for user actions, which may not cover all the nuances of stakeholder intentions, thus potentially introducing mono‐operation bias. We encourage future benchmark verification by external experts\rev{, similar to those carried out by, e.g., Chowdhury et al.~\cite{swebench-verified} for SWE-Bench~\cite{swebench}}. \rev{Additionally, the tasks we find in the literature are biased towards classification tasks, meaning our results might not generalize to other kinds of tasks that require SE language understanding. Furthermore, our fine-tuning protocol is restricted to LLMs of up to 7B parameters and we do not use repeated runs to select the best model. Both aspects might mean we are underestimating potential performances. However, we carefully configure the experiment setup and ensure stable convergence behavior for all experiments. Finally, our top-ranked LLM has 1.6B parameters, not being the largest one from those selected to run our experiments, which helps to mitigate these threats.}

\subsection{Internal Validity}


We apply a consistent pre-processing pipeline for most tasks, including (1) normalizing white spaces, (2) removing any kind of markdown and HTML styling, and (3) masking by some common regex patterns, such as URLs, hashes, user mentions, and code blocks; and use a stratified 80\%/20\% hold-out splitting strategy to control selection bias. All the fine-tuning experiments run on the same hardware, use the same libraries, and hyper-parameters (aside from differing epochs for NER and MLM). In addition, the proprietary LLMs are \rev{prompted via standard zero-shot and 3-shot prompting strategies with a common base template}; and for training the baselines, we follow a simplified but robust model selection process. In our replication package \cite{replicationpackage}, we publish the source code, datasets, prompt templates, and detailed results, to help others to identify and control further instrumentation threats. \rev{There is a risk that some LLMs may have seen some fine-tuning data during pre-training (e.g., sourced from GitHub or Stack Overflow), which might mean we overestimate performance with respect to unseen data.} 

\subsection{External Validity}


Considering that SELU is only focused on \rev{SE textual artifacts} NLU tasks and that the SE practice encompasses diverse contexts (application domains, programming languages, project types, etc.), our 22 tasks may not be fully representative. Even some of the most important progress on evaluating LLMs for solving code-related tasks (e.g., SWE-bench \cite{swebench}) fall into lack of diversity of sources and task types. In addition, although we manage to include tasks in most of the SE life cycle activities, \rev{7 task datasets derive from developer forums (e.g., Stack Overflow) and mailing lists (e.g., Linux Kernel)}, whose relevance declines as AI-driven workflows evolve. Future work should expand task diversity even more, include datasets from proprietary or industry settings, and validate results in live developer environments. \rev{Further, while the metrics we use are well-accepted for the evaluation of ML models, it is unclear to which degree they are representative for real-world SE value, possibly limiting how our conclusions can be generalized to system performance. This includes the consideration of whether human judgments should be replaced at all \cite{ahmed2025can}. Our benchmark also does not consider non-functional aspects of model performance. This includes ensuring that there are no harmful biases negatively affecting fairness~\cite{bano2025, morales2024}, but also a concrete consideration of costs (e.g., training costs, inference costs~\cite{sallou2924}). Finally, while benchmarks that aggregate performance across multiple tasks are well suited to understand general model capabilities, this does not imply that top-ranked ones are automatically the best in all covered tasks~\cite{reviews_prioritization}.}

\subsection{Conclusion Validity}


The conclusions that we derive in this work have some threats, including (i) limited statistical power from only \rev{20} paired classification and regression tasks, (ii) variability and potential bias in metrics (F1-macro, SMAPE, F1-micro, accuracy) despite their normalized range, and (iii) performance variance attributed to hyper-parameters and random seed choice~\cite{finetuning-stability}. We mitigate these threats by applying Shapiro–Wilk tests to check normality, using the Bayesian signed-rank tests with ROPE to distinguish meaningful differences, and reporting Bonferroni-corrected confidence intervals. Using early stopping and stratified splits might reduce performance variance, but single‐run fine-tuning and fixed prompts mean some results still depend on untested randomness and design decisions.

\section{Conclusions}
\label{s:conclusions}

In this paper, we introduce SELU, a comprehensive benchmark to evaluate LLMs on \rev{SE textual artifacts NLU} tasks throughout the SE life cycle. To the best of our knowledge, this is the first systematic effort to assemble such a diverse suite of tasks and to apply a Bayesian analysis for pairwise model comparison. We consider a variety of LLMs for evaluating them on SELU, including 22 open-source and two proprietary alternatives; while comparing them against two ML baselines. Open-source LLMs vary in model size and architecture, and they are generalist and domain-adapted \rev{via code-focused pre-training. Proprietary LLMs are prompted using zero-shot and 3-shot prompting strategies.}

Our results reveal \rev{three} key insights. First, decoder-only LLMs (GPT-2 and Llama 3.2 series) form a top performance tier, yielding the highest mean performance with relatively low across-task variance. \rev{However, other 12 LLMs are not far behind, including, e.g., the comparatively small encoder-only ModernBERT and BERT in their base versions.} Second, domain adaptation via code-focused pre-training \rev{does not yield significant improvements and can even hurt performance on developer communication tasks.} Finally, \rev{the best predictor for the expected performance is the task type, i.e., LLMs are likely to produce good performance for binary tasks but still struggle at more complex tasks such as multi-label classification.}

\section*{Acknowledgments}

This work is funded by the German Research Foundation (DFG) through the SENLP project under Grant 524228075.

\bibliographystyle{IEEEtran}
\bibliography{refs}



\appendices

\section{\rev{Task Inventory}}
\label{a:taskinventory}

In the following, \rev{we provide additional details for each task included in SELU: our rational for assigning it to a SE life cycle activity and task family and its class/label schema.}

\subsection{Binary classification tasks}

\subsubsection{\textit{bug\_issue}}

Classifying whether a reported issue is a bug is a core quality assurance activity because accurate bug labels uphold project quality gates (e.g., ensuring that all high-priority bugs are resolved before release) and prevent non-bug reports from unnecessarily blocking testing and deployment~\cite{bug_issue}. \rev{Although an issue initially requires a description in natural language, it may be accompanied by, e.g., an error log and the code block, or directly relate to a commit or pull request.}

\subsubsection{\rev{\textit{functional\_requirement}, \textit{quality\_requirement}, \textit{security\_ requirement}}}

\rev{Classifying requirements as having functional, quality, and security-related aspects are foundational (core non-code) requirements analysis tasks since they structure stakeholder needs, ensure that both behavioral and quality concerns are explicitly captured, and reduces downstream development risk~\cite{functional_requirement}. In particular, distinguishing security requirements early is critical to avoid late discovery of security gaps, which are costly to remediate and often stem from misclassified or overlooked requirements~\cite{security_requirement}.}

\subsubsection{\textit{incivility}}

Detecting uncivil language in developer communication enables community managers to flag and remove disrespectful comments that erode collaboration and contributor engagement, while maintaining a healthy and productive development  environment~\cite{incivility+tone_bearing1,incivility+tone_bearing2}.

\subsubsection{\rev{\textit{safety\_issue}}}

\rev{Identifying safety-related issues in code artifacts supports quality assurance by enabling early detection of defects or risky behaviors that may lead to system failures or hazards, which must be mitigated before release. In addition, it helps prioritizing verification and validation, reinforcing the quality assurance role in preventing harmful outcomes in safety-critical software systems~\cite{safety_issue}.}

\subsubsection{\textit{tone\_bearing}}

Detecting tone-bearing (non-technical) messages in developer communication, which `convey a mood or style of expression' such as frustration or excitement, enables community managers to surface emotionally charged discussions that may precede uncivil behavior and require closer moderation~\cite{incivility+tone_bearing1,incivility+tone_bearing2}.

\subsection{Multi-class classification tasks}

\subsubsection{\textit{closed\_question}}

Classifying whether a question should be closed assists moderators in managing \rev{developer forums} by flagging posts that are not real questions, off-topic, not constructive, or too localized; thus helping to maintain overall quality and uphold community standards~\cite{closed_question}.

\subsubsection{\textit{commit\_intention}}

Distinguishing commits as perfective, corrective, or other (Swanson’s categories) using developer‐stated intention in commit messages helps identify the nature of maintenance work. This classification supports the planning of targeted maintenance and the analysis of quality evolution~\cite{commit_intention}. \rev{Commit messages are closely related to changes in the codebase.}

\subsubsection{\rev{\textit{issue\_intention}}}

\rev{Classifying the intention of issues (e.g., aspect evaluation, feature request, information giving, information seeking, problem discovery, solution proposal, or others) directly supports software maintenance by enabling effective issue triage and prioritization in evolving software systems, while helps maintainers allocate effort and plan maintenance activities more efficiently across large issue trackers~\cite{issue_intention}. Although an issue initially requires a description in natural language, it may be accompanied by, e.g., an error log and the code block, or directly relate to a commit or pull request.}

\subsubsection{\textit{issue\_type}}

Classifying issues into bugs, enhancements, or questions directly supports software maintenance by streamlining issue triage and backlog prioritization. This task underpins effective issue management and resource allocation within maintenance workflows~\cite{issue_type}. \rev{Although an issue initially requires a description in natural language, it may be accompanied by, e.g., an error log and the code block that originates it, or directly relate to a commit or pull request.}

\subsubsection{\textit{question\_quality}}

Question quality classification \rev{supports content moderation management} by flagging high versus low-quality posts, enabling subsequent prioritization of editing or closure actions. This helps maintain high standards in developer communications~\cite{question_quality}.

\subsubsection{\rev{\textit{review\_type}}}

\rev{Classifying app reviews by type (e.g., information giving, information seeking, feature request, or problem discovery) supports software maintenance by enabling developers to efficiently triage large volumes of user feedback and focus on actionable inputs for software evolution. Distinguishing review types helps prioritize corrective and adaptive maintenance activities by filtering noise and surfacing reviews relevant to fixing defects or planning future releases~\cite{review_type}.}

\subsubsection{\textit{sentiment}}

Analyzing sentiments (positive, negative, neutral) equips managers with quantifiable insights into attitudes of developers and users (e.g., frustration, satisfaction), enabling monitoring of team morale and collaboration health. Emotions can influence work outcomes such as productivity, task quality, and group harmony, making sentiment metrics valuable for proactive management~\cite{sentiment}.

\subsection{Multi-label classification tasks}

\subsubsection{\textit{comment\_type\_java}, \textit{comment\_type\_pharo}, \textit{comment\_ type\_python}}

Structuring comments into semantic information types helps developers locate the precise documentation they need while understanding and writing code. By applying the Class Comment Type Model (CCTM) taxonomy, comment classification directly supports developers at all stages of implementation by tailoring comment content to their immediate development tasks~\cite{comment_type}. \rev{Comment types across languages are slightly different. Summary, ownership, expand, usage, pointer, deprecation, and rationale, for Java. Key implementation points, example, responsibilities, class references, intention, key messages, and collaborators, for Pharo. Usage, parameters, development notes, expand, and summary, for Python.}

\subsubsection{\textit{review\_aspect}}

Categorizing API reviews into relevant aspects supports quality assurance. It operationalizes aspect‐focused concerns into actionable quality indicators, enabling systematic evaluation and continuous improvement of the software throughout the SE life cycle~\cite{review_aspect}. \rev{The aspects are varied: bug, community, compatibility, documentation, legal, only sentiment, performance, portability, security, usability, and others.}

\subsubsection{\textit{smell\_doc}}

API documentation smells are presentation‐level quality defects (analogous to code smells) that, while not causing functional errors, prevent software/code learnability and reuse, making their detection a core responsibility of quality assurance~\cite{smell_doc}. \rev{Documentation smells include: bloated, excessive, fragmented, lazy, and tangled.}

\subsection{Regression tasks}

\subsubsection{\textit{story\_points}}

Story point estimation quantifies the effort and complexity of backlog items, enabling managers to plan sprint capacity and allocate resources effectively while avoiding idle time or cost overruns~\cite{story_points}.

\begin{figure*}[t]
\centering
\caption{\rev{Posterior probability matrices of (a) `A outperforming B,' (b) `B outperforming A,' and (c) `A and B being practically equivalent.' (d) Represents inconclusive result (i.e., none of the probabilities is higher than $0.95$).}}
\label{fig:posterior_matrices}
\includegraphics[width=0.84\textwidth]{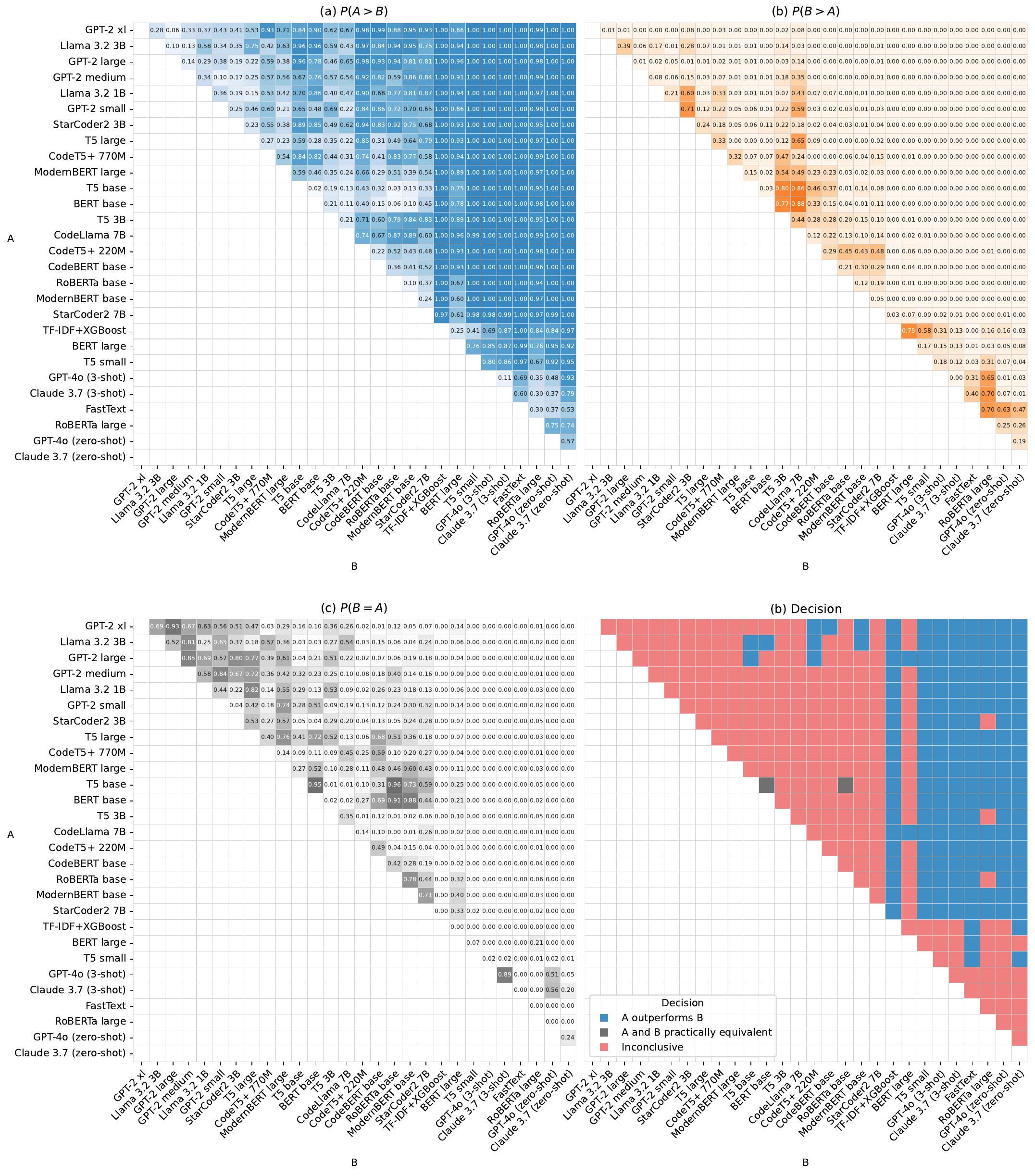}
\end{figure*}

\subsection{NER tasks}

\subsubsection{\textit{se\_entities}}

\rev{Recognizing SE-related terminology (entities) from developer forums directly supports context comprehension of discussions, questions, or answers; aiding developers during any stage of the implementation process~\cite{se_entities}. Among the most representative entities are class, application, variable, user interface element, code block, function, and library.}

\subsection{MLM tasks}

\subsubsection{\textit{requirement\_completion}}

Verifying the completion of requirements entails ensuring that each requirement fully specifies its intended functionality and quality constraints, addressing both internal and external notions of completeness. A precise definition of a requirement is crucial because omissions in the specification can lead to misinterpretation, latent defects, and costly rework~\cite{requirement_completion}.







\section{\rev{Additional details for the results.}}
\label{a:results-details}

Performance scores based on F1-macro, SMAPE, F1-micro, and accuracy for classification, regression, NER and MLM tasks, respectively, are detailed in Table~\ref{tbl:results1}. Green cells highlight the outperforming model for each task.

Figure \ref{fig:posterior_matrices} shows the posterior probabilities, $\mathbb{P}(A > B)$, $\mathbb{P}(A = B)$, and $\mathbb{P}(B > A)$, for `A outperforming B,' `A and B being practically equivalent,' and `B outperforming A,' respectively. We conclude that $A$ outperforms $B$ if $\mathbb{P}(A > B) \ge 0.95$, that they are equivalent if $\mathbb{P}(A = B) \ge 0.95$, and that $B$ outperforms $A$ if $\mathbb{P}(B > A) \ge 0.95$; otherwise, the result is inconclusive.





\begin{sidewaystable*}[t]
\centering
\caption{\rev{Evaluation results for all tasks. Results aggregated by task family do not include NER and MLM tasks, since they are not available for all models.}}
\label{tbl:results1}
\resizebox{\textwidth}{!}{%
\begin{tblr}{
  column{1} = {c},
  cell{1}{2} = {c,b},
  cell{2}{1} = {r=7}{},
  cell{2}{3} = {c},
  cell{2}{4} = {c},
  cell{2}{5} = {c},
  cell{2}{6} = {c},
  cell{2}{7} = {c},
  cell{2}{8} = {c},
  cell{2}{9} = {c},
  cell{2}{10} = {c},
  cell{2}{11} = {c},
  cell{2}{12} = {c},
  cell{2}{13} = {c},
  cell{2}{14} = {c},
  cell{2}{15} = {c},
  cell{2}{16} = {c},
  cell{2}{17} = {c},
  cell{2}{18} = {c},
  cell{2}{19} = {c},
  cell{2}{20} = {green,c},
  cell{2}{21} = {c},
  cell{2}{22} = {c},
  cell{2}{23} = {c},
  cell{2}{24} = {c},
  cell{2}{25} = {c},
  cell{2}{26} = {c},
  cell{2}{27} = {c},
  cell{2}{28} = {c},
  cell{2}{29} = {c},
  cell{2}{30} = {c},
  cell{3}{3} = {c},
  cell{3}{4} = {c},
  cell{3}{5} = {c},
  cell{3}{6} = {c},
  cell{3}{7} = {c},
  cell{3}{8} = {c},
  cell{3}{9} = {c},
  cell{3}{10} = {green,c},
  cell{3}{11} = {c},
  cell{3}{12} = {c},
  cell{3}{13} = {c},
  cell{3}{14} = {c},
  cell{3}{15} = {c},
  cell{3}{16} = {c},
  cell{3}{17} = {c},
  cell{3}{18} = {c},
  cell{3}{19} = {c},
  cell{3}{20} = {c},
  cell{3}{21} = {c},
  cell{3}{22} = {c},
  cell{3}{23} = {c},
  cell{3}{24} = {c},
  cell{3}{25} = {c},
  cell{3}{26} = {c},
  cell{3}{27} = {c},
  cell{3}{28} = {c},
  cell{3}{29} = {c},
  cell{3}{30} = {c},
  cell{4}{3} = {c},
  cell{4}{4} = {c},
  cell{4}{5} = {c},
  cell{4}{6} = {c},
  cell{4}{7} = {c},
  cell{4}{8} = {c},
  cell{4}{9} = {c},
  cell{4}{10} = {c},
  cell{4}{11} = {green,c},
  cell{4}{12} = {c},
  cell{4}{13} = {c},
  cell{4}{14} = {c},
  cell{4}{15} = {c},
  cell{4}{16} = {c},
  cell{4}{17} = {c},
  cell{4}{18} = {c},
  cell{4}{19} = {c},
  cell{4}{20} = {c},
  cell{4}{21} = {c},
  cell{4}{22} = {c},
  cell{4}{23} = {c},
  cell{4}{24} = {c},
  cell{4}{25} = {c},
  cell{4}{26} = {c},
  cell{4}{27} = {c},
  cell{4}{28} = {c},
  cell{4}{29} = {c},
  cell{4}{30} = {c},
  cell{5}{3} = {c},
  cell{5}{4} = {c},
  cell{5}{5} = {c},
  cell{5}{6} = {c},
  cell{5}{7} = {c},
  cell{5}{8} = {c},
  cell{5}{9} = {c},
  cell{5}{10} = {c},
  cell{5}{11} = {c},
  cell{5}{12} = {c},
  cell{5}{13} = {c},
  cell{5}{14} = {c},
  cell{5}{15} = {c},
  cell{5}{16} = {c},
  cell{5}{17} = {green,c},
  cell{5}{18} = {c},
  cell{5}{19} = {c},
  cell{5}{20} = {c},
  cell{5}{21} = {c},
  cell{5}{22} = {c},
  cell{5}{23} = {c},
  cell{5}{24} = {c},
  cell{5}{25} = {c},
  cell{5}{26} = {c},
  cell{5}{27} = {c},
  cell{5}{28} = {c},
  cell{5}{29} = {c},
  cell{5}{30} = {c},
  cell{6}{3} = {c},
  cell{6}{4} = {c},
  cell{6}{5} = {c},
  cell{6}{6} = {c},
  cell{6}{7} = {green,c},
  cell{6}{8} = {c},
  cell{6}{9} = {c},
  cell{6}{10} = {c},
  cell{6}{11} = {c},
  cell{6}{12} = {c},
  cell{6}{13} = {c},
  cell{6}{14} = {c},
  cell{6}{15} = {c},
  cell{6}{16} = {c},
  cell{6}{17} = {c},
  cell{6}{18} = {c},
  cell{6}{19} = {c},
  cell{6}{20} = {c},
  cell{6}{21} = {c},
  cell{6}{22} = {c},
  cell{6}{23} = {c},
  cell{6}{24} = {c},
  cell{6}{25} = {c},
  cell{6}{26} = {c},
  cell{6}{27} = {c},
  cell{6}{28} = {c},
  cell{6}{29} = {c},
  cell{6}{30} = {c},
  cell{7}{3} = {c},
  cell{7}{4} = {c},
  cell{7}{5} = {c},
  cell{7}{6} = {c},
  cell{7}{7} = {c},
  cell{7}{8} = {c},
  cell{7}{9} = {c},
  cell{7}{10} = {c},
  cell{7}{11} = {c},
  cell{7}{12} = {c},
  cell{7}{13} = {c},
  cell{7}{14} = {c},
  cell{7}{15} = {c},
  cell{7}{16} = {c},
  cell{7}{17} = {c},
  cell{7}{18} = {c},
  cell{7}{19} = {c},
  cell{7}{20} = {c},
  cell{7}{21} = {c},
  cell{7}{22} = {c},
  cell{7}{23} = {c},
  cell{7}{24} = {green,c},
  cell{7}{25} = {c},
  cell{7}{26} = {c},
  cell{7}{27} = {c},
  cell{7}{28} = {c},
  cell{7}{29} = {c},
  cell{7}{30} = {c},
  cell{8}{3} = {c},
  cell{8}{4} = {c},
  cell{8}{5} = {c},
  cell{8}{6} = {c},
  cell{8}{7} = {c},
  cell{8}{8} = {c},
  cell{8}{9} = {c},
  cell{8}{10} = {c},
  cell{8}{11} = {c},
  cell{8}{12} = {c},
  cell{8}{13} = {c},
  cell{8}{14} = {green,c},
  cell{8}{15} = {c},
  cell{8}{16} = {c},
  cell{8}{17} = {c},
  cell{8}{18} = {c},
  cell{8}{19} = {c},
  cell{8}{20} = {c},
  cell{8}{21} = {c},
  cell{8}{22} = {c},
  cell{8}{23} = {c},
  cell{8}{24} = {c},
  cell{8}{25} = {c},
  cell{8}{26} = {c},
  cell{8}{27} = {c},
  cell{8}{28} = {c},
  cell{8}{29} = {c},
  cell{8}{30} = {c},
  cell{9}{1} = {r=7}{},
  cell{9}{3} = {c},
  cell{9}{4} = {c},
  cell{9}{5} = {c},
  cell{9}{6} = {c},
  cell{9}{7} = {c},
  cell{9}{8} = {c},
  cell{9}{9} = {c},
  cell{9}{10} = {c},
  cell{9}{11} = {c},
  cell{9}{12} = {c},
  cell{9}{13} = {c},
  cell{9}{14} = {c},
  cell{9}{15} = {c},
  cell{9}{16} = {c},
  cell{9}{17} = {c},
  cell{9}{18} = {c},
  cell{9}{19} = {c},
  cell{9}{20} = {green,c},
  cell{9}{21} = {c},
  cell{9}{22} = {c},
  cell{9}{23} = {c},
  cell{9}{24} = {c},
  cell{9}{25} = {c},
  cell{9}{26} = {c},
  cell{9}{27} = {c},
  cell{9}{28} = {c},
  cell{9}{29} = {c},
  cell{9}{30} = {c},
  cell{10}{3} = {c},
  cell{10}{4} = {c},
  cell{10}{5} = {c},
  cell{10}{6} = {c},
  cell{10}{7} = {c},
  cell{10}{8} = {c},
  cell{10}{9} = {c},
  cell{10}{10} = {c},
  cell{10}{11} = {c},
  cell{10}{12} = {c},
  cell{10}{13} = {c},
  cell{10}{14} = {c},
  cell{10}{15} = {c},
  cell{10}{16} = {c},
  cell{10}{17} = {c},
  cell{10}{18} = {c},
  cell{10}{19} = {c},
  cell{10}{20} = {c},
  cell{10}{21} = {green,c},
  cell{10}{22} = {c},
  cell{10}{23} = {c},
  cell{10}{24} = {c},
  cell{10}{25} = {c},
  cell{10}{26} = {c},
  cell{10}{27} = {c},
  cell{10}{28} = {c},
  cell{10}{29} = {c},
  cell{10}{30} = {c},
  cell{11}{3} = {c},
  cell{11}{4} = {c},
  cell{11}{5} = {c},
  cell{11}{6} = {c},
  cell{11}{7} = {c},
  cell{11}{8} = {c},
  cell{11}{9} = {c},
  cell{11}{10} = {c},
  cell{11}{11} = {c},
  cell{11}{12} = {c},
  cell{11}{13} = {c},
  cell{11}{14} = {c},
  cell{11}{15} = {c},
  cell{11}{16} = {c},
  cell{11}{17} = {c},
  cell{11}{18} = {c},
  cell{11}{19} = {c},
  cell{11}{20} = {c},
  cell{11}{21} = {c},
  cell{11}{22} = {green,c},
  cell{11}{23} = {c},
  cell{11}{24} = {c},
  cell{11}{25} = {c},
  cell{11}{26} = {c},
  cell{11}{27} = {c},
  cell{11}{28} = {c},
  cell{11}{29} = {c},
  cell{11}{30} = {c},
  cell{12}{3} = {c},
  cell{12}{4} = {c},
  cell{12}{5} = {c},
  cell{12}{6} = {c},
  cell{12}{7} = {c},
  cell{12}{8} = {c},
  cell{12}{9} = {c},
  cell{12}{10} = {c},
  cell{12}{11} = {c},
  cell{12}{12} = {c},
  cell{12}{13} = {c},
  cell{12}{14} = {c},
  cell{12}{15} = {c},
  cell{12}{16} = {c},
  cell{12}{17} = {c},
  cell{12}{18} = {c},
  cell{12}{19} = {c},
  cell{12}{20} = {green,c},
  cell{12}{21} = {c},
  cell{12}{22} = {c},
  cell{12}{23} = {c},
  cell{12}{24} = {c},
  cell{12}{25} = {c},
  cell{12}{26} = {c},
  cell{12}{27} = {c},
  cell{12}{28} = {c},
  cell{12}{29} = {c},
  cell{12}{30} = {c},
  cell{13}{3} = {c},
  cell{13}{4} = {c},
  cell{13}{5} = {c},
  cell{13}{6} = {c},
  cell{13}{7} = {c},
  cell{13}{8} = {c},
  cell{13}{9} = {c},
  cell{13}{10} = {c},
  cell{13}{11} = {c},
  cell{13}{12} = {c},
  cell{13}{13} = {c},
  cell{13}{14} = {c},
  cell{13}{15} = {c},
  cell{13}{16} = {c},
  cell{13}{17} = {c},
  cell{13}{18} = {c},
  cell{13}{19} = {c},
  cell{13}{20} = {green,c},
  cell{13}{21} = {c},
  cell{13}{22} = {c},
  cell{13}{23} = {c},
  cell{13}{24} = {c},
  cell{13}{25} = {c},
  cell{13}{26} = {c},
  cell{13}{27} = {c},
  cell{13}{28} = {c},
  cell{13}{29} = {c},
  cell{13}{30} = {c},
  cell{14}{3} = {c},
  cell{14}{4} = {c},
  cell{14}{5} = {c},
  cell{14}{6} = {c},
  cell{14}{7} = {c},
  cell{14}{8} = {c},
  cell{14}{9} = {c},
  cell{14}{10} = {c},
  cell{14}{11} = {c},
  cell{14}{12} = {green,c},
  cell{14}{13} = {c},
  cell{14}{14} = {c},
  cell{14}{15} = {c},
  cell{14}{16} = {c},
  cell{14}{17} = {c},
  cell{14}{18} = {c},
  cell{14}{19} = {c},
  cell{14}{20} = {c},
  cell{14}{21} = {c},
  cell{14}{22} = {c},
  cell{14}{23} = {c},
  cell{14}{24} = {c},
  cell{14}{25} = {c},
  cell{14}{26} = {c},
  cell{14}{27} = {c},
  cell{14}{28} = {c},
  cell{14}{29} = {c},
  cell{14}{30} = {c},
  cell{15}{3} = {c},
  cell{15}{4} = {c},
  cell{15}{5} = {c},
  cell{15}{6} = {c},
  cell{15}{7} = {c},
  cell{15}{8} = {c},
  cell{15}{9} = {c},
  cell{15}{10} = {c},
  cell{15}{11} = {c},
  cell{15}{12} = {c},
  cell{15}{13} = {c},
  cell{15}{14} = {c},
  cell{15}{15} = {c},
  cell{15}{16} = {c},
  cell{15}{17} = {c},
  cell{15}{18} = {green,c},
  cell{15}{19} = {c},
  cell{15}{20} = {c},
  cell{15}{21} = {c},
  cell{15}{22} = {c},
  cell{15}{23} = {c},
  cell{15}{24} = {c},
  cell{15}{25} = {c},
  cell{15}{26} = {c},
  cell{15}{27} = {c},
  cell{15}{28} = {c},
  cell{15}{29} = {c},
  cell{15}{30} = {c},
  cell{16}{1} = {r=5}{},
  cell{16}{3} = {c},
  cell{16}{4} = {c},
  cell{16}{5} = {c},
  cell{16}{6} = {c},
  cell{16}{7} = {c},
  cell{16}{8} = {c},
  cell{16}{9} = {c},
  cell{16}{10} = {c},
  cell{16}{11} = {c},
  cell{16}{12} = {c},
  cell{16}{13} = {c},
  cell{16}{14} = {c},
  cell{16}{15} = {c},
  cell{16}{16} = {c},
  cell{16}{17} = {c},
  cell{16}{18} = {c},
  cell{16}{19} = {c},
  cell{16}{20} = {c},
  cell{16}{21} = {c},
  cell{16}{22} = {c},
  cell{16}{23} = {green,c},
  cell{16}{24} = {c},
  cell{16}{25} = {c},
  cell{16}{26} = {c},
  cell{16}{27} = {c},
  cell{16}{28} = {c},
  cell{16}{29} = {c},
  cell{16}{30} = {c},
  cell{17}{3} = {c},
  cell{17}{4} = {c},
  cell{17}{5} = {c},
  cell{17}{6} = {c},
  cell{17}{7} = {c},
  cell{17}{8} = {c},
  cell{17}{9} = {green,c},
  cell{17}{10} = {c},
  cell{17}{11} = {c},
  cell{17}{12} = {c},
  cell{17}{13} = {c},
  cell{17}{14} = {c},
  cell{17}{15} = {c},
  cell{17}{16} = {c},
  cell{17}{17} = {c},
  cell{17}{18} = {c},
  cell{17}{19} = {c},
  cell{17}{20} = {c},
  cell{17}{21} = {c},
  cell{17}{22} = {c},
  cell{17}{23} = {c},
  cell{17}{24} = {c},
  cell{17}{25} = {c},
  cell{17}{26} = {c},
  cell{17}{27} = {c},
  cell{17}{28} = {c},
  cell{17}{29} = {c},
  cell{17}{30} = {c},
  cell{18}{3} = {c},
  cell{18}{4} = {c},
  cell{18}{5} = {c},
  cell{18}{6} = {c},
  cell{18}{7} = {c},
  cell{18}{8} = {c},
  cell{18}{9} = {c},
  cell{18}{10} = {c},
  cell{18}{11} = {green,c},
  cell{18}{12} = {c},
  cell{18}{13} = {c},
  cell{18}{14} = {c},
  cell{18}{15} = {c},
  cell{18}{16} = {c},
  cell{18}{17} = {c},
  cell{18}{18} = {c},
  cell{18}{19} = {c},
  cell{18}{20} = {c},
  cell{18}{21} = {c},
  cell{18}{22} = {c},
  cell{18}{23} = {c},
  cell{18}{24} = {c},
  cell{18}{25} = {c},
  cell{18}{26} = {c},
  cell{18}{27} = {c},
  cell{18}{28} = {c},
  cell{18}{29} = {c},
  cell{18}{30} = {c},
  cell{19}{3} = {c},
  cell{19}{4} = {c},
  cell{19}{5} = {c},
  cell{19}{6} = {c},
  cell{19}{7} = {c},
  cell{19}{8} = {c},
  cell{19}{9} = {c},
  cell{19}{10} = {c},
  cell{19}{11} = {c},
  cell{19}{12} = {c},
  cell{19}{13} = {c},
  cell{19}{14} = {c},
  cell{19}{15} = {c},
  cell{19}{16} = {c},
  cell{19}{17} = {c},
  cell{19}{18} = {c},
  cell{19}{19} = {c},
  cell{19}{20} = {c},
  cell{19}{21} = {green,c},
  cell{19}{22} = {c},
  cell{19}{23} = {c},
  cell{19}{24} = {c},
  cell{19}{25} = {c},
  cell{19}{26} = {c},
  cell{19}{27} = {c},
  cell{19}{28} = {c},
  cell{19}{29} = {c},
  cell{19}{30} = {c},
  cell{20}{3} = {c},
  cell{20}{4} = {c},
  cell{20}{5} = {c},
  cell{20}{6} = {c},
  cell{20}{7} = {c},
  cell{20}{8} = {c},
  cell{20}{9} = {c},
  cell{20}{10} = {c},
  cell{20}{11} = {c},
  cell{20}{12} = {c},
  cell{20}{13} = {c},
  cell{20}{14} = {c},
  cell{20}{15} = {c},
  cell{20}{16} = {c},
  cell{20}{17} = {c},
  cell{20}{18} = {c},
  cell{20}{19} = {c},
  cell{20}{20} = {c},
  cell{20}{21} = {c},
  cell{20}{22} = {c},
  cell{20}{23} = {c},
  cell{20}{24} = {c},
  cell{20}{25} = {c},
  cell{20}{26} = {c},
  cell{20}{27} = {c},
  cell{20}{28} = {c},
  cell{20}{29} = {green,c},
  cell{20}{30} = {c},
  cell{21}{3} = {c},
  cell{21}{4} = {c},
  cell{21}{5} = {c},
  cell{21}{6} = {c},
  cell{21}{7} = {c},
  cell{21}{8} = {c},
  cell{21}{9} = {c},
  cell{21}{10} = {c},
  cell{21}{11} = {c},
  cell{21}{12} = {c},
  cell{21}{13} = {c},
  cell{21}{14} = {c},
  cell{21}{15} = {c},
  cell{21}{16} = {c},
  cell{21}{17} = {c},
  cell{21}{18} = {c},
  cell{21}{19} = {c},
  cell{21}{20} = {c},
  cell{21}{21} = {c},
  cell{21}{22} = {c},
  cell{21}{23} = {c},
  cell{21}{24} = {green,c},
  cell{21}{25} = {c},
  cell{21}{26} = {c},
  cell{21}{27} = {c},
  cell{21}{28} = {c},
  cell{21}{29} = {c},
  cell{21}{30} = {c},
  cell{22}{3} = {c},
  cell{22}{4} = {c},
  cell{22}{5} = {c},
  cell{22}{6} = {green,c},
  cell{22}{7} = {c},
  cell{22}{8} = {c},
  cell{22}{9} = {c},
  cell{22}{10} = {c},
  cell{22}{11} = {c},
  cell{22}{12} = {c},
  cell{22}{13} = {c},
  cell{22}{14} = {c},
  cell{22}{15} = {c},
  cell{22}{16} = {c},
  cell{22}{17} = {c},
  cell{22}{18} = {c},
  cell{22}{19} = {c},
  cell{22}{20} = {c},
  cell{22}{21} = {c},
  cell{22}{22} = {c},
  cell{22}{23} = {c},
  cell{22}{24} = {c},
  cell{22}{25} = {c},
  cell{22}{26} = {c},
  cell{22}{27} = {c},
  cell{22}{28} = {c},
  cell{22}{29} = {c},
  cell{22}{30} = {c},
  cell{23}{3} = {c},
  cell{23}{4} = {c},
  cell{23}{5} = {c},
  cell{23}{6} = {c},
  cell{23}{7} = {c},
  cell{23}{8} = {green,c},
  cell{23}{9} = {c},
  cell{23}{10} = {c},
  cell{23}{11} = {c},
  cell{23}{12} = {c},
  cell{23}{13} = {c},
  cell{23}{14} = {c},
  cell{23}{15} = {c},
  cell{23}{16} = {c},
  cell{23}{17} = {c},
  cell{23}{18} = {c},
  cell{23}{19} = {c},
  cell{23}{20} = {c},
  cell{23}{21} = {c},
  cell{23}{22} = {c},
  cell{23}{23} = {c},
  cell{23}{24} = {c},
  cell{23}{25} = {c},
  cell{23}{26} = {c},
  cell{23}{27} = {c},
  cell{23}{28} = {c},
  cell{23}{29} = {c},
  cell{23}{30} = {c},
  cell{24}{1} = {r=3}{},
  cell{24}{3} = {c=16}{c},
  cell{24}{19} = {c=6}{c},
  cell{24}{25} = {c=4}{c},
  cell{24}{29} = {c=2}{c},
  cell{25}{3} = {c=16}{c},
  cell{25}{19} = {c=6}{c},
  cell{25}{25} = {c=4}{c},
  cell{25}{29} = {c=2}{c},
  cell{26}{3} = {c=16}{c},
  cell{26}{19} = {c=6}{c},
  cell{26}{25} = {c=4}{c},
  cell{26}{29} = {c=2}{c},
  cell{27}{1} = {r=6}{},
  cell{27}{3} = {c=16}{c},
  cell{27}{19} = {c=6}{c},
  cell{27}{25} = {c=4}{c},
  cell{27}{29} = {c=2}{c},
  cell{28}{3} = {c=16}{c},
  cell{28}{19} = {c=6}{c},
  cell{28}{25} = {c=4}{c},
  cell{28}{29} = {c=2}{c},
  cell{29}{3} = {c=16}{c},
  cell{29}{19} = {c=6}{c},
  cell{29}{25} = {c=4}{c},
  cell{29}{29} = {c=2}{c},
  cell{30}{3} = {c=16}{c},
  cell{30}{19} = {c=6}{c},
  cell{30}{25} = {c=4}{c},
  cell{30}{29} = {c=2}{c},
  cell{31}{3} = {c=16}{c},
  cell{31}{19} = {c=6}{c},
  cell{31}{25} = {c=4}{c},
  cell{31}{29} = {c=2}{c},
  cell{32}{3} = {c=16}{c},
  cell{32}{19} = {c=6}{c},
  cell{32}{25} = {c=4}{c},
  cell{32}{29} = {c=2}{c},
  vline{3,19,25,29} = {1-32}{},
  hline{1,33} = {-}{0.11em},
  hline{2,9,16,21} = {-}{},
  hline{24,27} = {-}{wd=0.11em},
}
 & \textbf{Task ID} & \begin{sideways}\textbf{BERT base}\end{sideways} & \begin{sideways}\textbf{BERT large}\end{sideways} & \begin{sideways}\textbf{RoBERTa base}\end{sideways} & \begin{sideways}\textbf{RoBERTa large}\end{sideways} & \begin{sideways}\textbf{ModernBERT base}\end{sideways} & \begin{sideways}\textbf{ModernBERT large}\end{sideways} & \begin{sideways}\textbf{GPT-2 small}\end{sideways} & \begin{sideways}\textbf{GPT-2 medium}\end{sideways} & \begin{sideways}\textbf{GPT-2 large}\end{sideways} & \begin{sideways}\textbf{GPT-2 xl}\end{sideways} & \begin{sideways}\textbf{Llama 3.2 1B}\end{sideways} & \begin{sideways}\textbf{Llama 3.2 3B}\end{sideways} & \begin{sideways}\textbf{T5 small}\end{sideways} & \begin{sideways}\textbf{T5 base}\end{sideways} & \begin{sideways}\textbf{T5 large}\end{sideways} & \begin{sideways}\textbf{T5 3B}\end{sideways} & \begin{sideways}\textbf{CodeBERT base}\end{sideways} & \begin{sideways}\textbf{CodeLlama 7B}\end{sideways} & \begin{sideways}\textbf{StarCoder2 3B}\end{sideways} & \begin{sideways}\textbf{StarCoder2 7B}\end{sideways} & \begin{sideways}\textbf{CodeT5+ 220M}\end{sideways} & \begin{sideways}\textbf{CodeT5+ 770M}\end{sideways} & \begin{sideways}\textbf{GPT-4o (zero-shot)}\end{sideways} & \begin{sideways}\textbf{Claude 3.7 (zero-shot)}\end{sideways} & \begin{sideways}\textbf{GPT-4o (3-shot)}\end{sideways} & \begin{sideways}\textbf{Claude 3.7 (3-shot)}\end{sideways} & \begin{sideways}\textbf{TF-IDF+XGBoost}\end{sideways} & \begin{sideways}\textbf{FastText}\end{sideways}\\
\begin{sideways}Binary\end{sideways} & \textit{bug\_issue} & 0.864 & 0.873 & 0.873 & 0.883 & 0.867 & 0.884 & 0.866 & 0.876 & 0.871 & 0.88 & 0.862 & 0.872 & 0.832 & 0.866 & 0.877 & 0.884 & 0.875 & 0.885 & 0.883 & 0.88 & 0.872 & 0.863 & 0.823 & 0.804 & 0.817 & 0.806 & 0.776 & 0.798\\
 & \textit{functional\_requirement} & 0.843 & 0.843 & 0.739 & 0.856 & 0.843 & 0.82 & 0.82 & 0.874 & 0.83 & 0.844 & 0.835 & 0.858 & 0.804 & 0.869 & 0.84 & 0.855 & 0.814 & 0.834 & 0.849 & 0.851 & 0.837 & 0.853 & 0.798 & 0.852 & 0.768 & 0.811 & 0.719 & 0.676\\
 & \textit{incivility} & 0.751 & 0.758 & 0.771 & 0.411 & 0.74 & 0.724 & 0.738 & 0.768 & 0.815 & 0.784 & 0.791 & 0.747 & 0.725 & 0.778 & 0.801 & 0.809 & 0.689 & 0.555 & 0.775 & 0.747 & 0.669 & 0.646 & 0.762 & 0.652 & 0.759 & 0.664 & 0.634 & 0.595\\
 & \textit{quality\_requirement} & 0.834 & 0.854 & 0.802 & 0.84 & 0.834 & 0.828 & 0.849 & 0.844 & 0.844 & 0.851 & 0.846 & 0.85 & 0.774 & 0.796 & 0.864 & 0.846 & 0.811 & 0.4 & 0.815 & 0.797 & 0.821 & 0.787 & 0.733 & 0.813 & 0.806 & 0.818 & 0.78 & 0.722\\
 & \textit{safety\_issue} & 0.854 & 0.86 & 0.85 & 0.845 & 0.876 & 0.851 & 0.856 & 0.863 & 0.84 & 0.843 & 0.832 & 0.818 & 0.85 & 0.838 & 0.855 & 0.83 & 0.849 & 0.838 & 0.862 & 0.864 & 0.784 & 0.847 & 0.791 & 0.694 & 0.815 & 0.777 & 0.788 & 0.774\\
 & \textit{security\_requirement} & 0.936 & 0.389 & 0.915 & 0.389 & 0.908 & 0.936 & 0.947 & 0.913 & 0.947 & 0.935 & 0.885 & 0.948 & 0.715 & 0.916 & 0.958 & 0.968 & 0.958 & 0.968 & 0.948 & 0.767 & 0.937 & 0.979 & 0.887 & 0.783 & 0.907 & 0.869 & 0.773 & 0.389\\
 & \textit{tone\_bearing} & 0.65 & 0.709 & 0.68 & 0.465 & 0.566 & 0.681 & 0.686 & 0.57 & 0.563 & 0.714 & 0.723 & 0.741 & 0.688 & 0.7 & 0.642 & 0.734 & 0.465 & 0.706 & 0.553 & 0.703 & 0.471 & 0.588 & 0.447 & 0.431 & 0.473 & 0.422 & 0.646 & 0.641\\
\begin{sideways}Multi-class\end{sideways} & \textit{closed\_question} & 0.533 & 0.555 & 0.55 & 0.569 & 0.551 & 0.567 & 0.548 & 0.541 & 0.547 & 0.553 & 0.558 & 0.549 & 0.5 & 0.551 & 0.564 & 0.563 & 0.546 & 0.59 & 0.585 & 0.561 & 0.544 & 0.537 & 0.381 & 0.412 & 0.4 & 0.397 & 0.467 & 0.477\\
 & \textit{commit\_intention} & 0.719 & 0.642 & 0.75 & 0.192 & 0.715 & 0.758 & 0.714 & 0.751 & 0.737 & 0.73 & 0.743 & 0.756 & 0.68 & 0.718 & 0.742 & 0.564 & 0.77 & 0.706 & 0.78 & 0.694 & 0.71 & 0.765 & 0.691 & 0.665 & 0.734 & 0.694 & 0.666 & 0.605\\
 & \textit{issue\_intention} & 0.843 & 0.85 & 0.866 & 0.855 & 0.848 & 0.861 & 0.848 & 0.86 & 0.873 & 0.85 & 0.861 & 0.843 & 0.859 & 0.847 & 0.859 & 0.879 & 0.856 & 0.845 & 0.871 & 0.882 & 0.815 & 0.834 & 0.763 & 0.645 & 0.777 & 0.705 & 0.736 & 0.687\\
 & \textit{issue\_type} & 0.782 & 0.786 & 0.792 & 0.798 & 0.793 & 0.803 & 0.792 & 0.79 & 0.794 & 0.797 & 0.796 & 0.802 & 0.742 & 0.774 & 0.786 & 0.8 & 0.791 & 0.814 & 0.808 & 0.807 & 0.763 & 0.773 & 0.758 & 0.729 & 0.747 & 0.74 & 0.714 & 0.723\\
 & \textit{question\_quality} & 0.848 & 0.841 & 0.857 & 0.866 & 0.862 & 0.876 & 0.862 & 0.869 & 0.874 & 0.872 & 0.87 & 0.867 & 0.819 & 0.852 & 0.86 & 0.868 & 0.854 & 0.881 & 0.861 & 0.855 & 0.862 & 0.869 & 0.339 & 0.346 & 0.434 & 0.411 & 0.764 & 0.795\\
 & \textit{review\_type} & 0.786 & 0.392 & 0.775 & 0.747 & 0.761 & 0.724 & 0.587 & 0.789 & 0.778 & 0.812 & 0.757 & 0.713 & 0.756 & 0.789 & 0.766 & 0.454 & 0.773 & 0.679 & 0.735 & 0.773 & 0.707 & 0.76 & 0.767 & 0.756 & 0.748 & 0.767 & 0.641 & 0.463\\
 & \textit{sentiment} & 0.803 & 0.801 & 0.802 & 0.806 & 0.8 & 0.797 & 0.799 & 0.793 & 0.811 & 0.808 & 0.812 & 0.777 & 0.736 & 0.8 & 0.808 & 0.815 & 0.782 & 0.8 & 0.806 & 0.752 & 0.771 & 0.786 & 0.684 & 0.712 & 0.689 & 0.725 & 0.626 & 0.588\\
\begin{sideways}Multi-label\end{sideways} & \textit{comment\_type\_java} & 0.768 & 0.789 & 0.772 & 0.643 & 0.772 & 0.812 & 0.785 & 0.798 & 0.795 & 0.796 & 0.727 & 0.809 & 0.594 & 0.77 & 0.787 & 0.662 & 0.718 & 0.823 & 0.804 & 0.792 & 0.824 & 0.807 & 0.293 & 0.429 & 0.402 & 0.466 & 0.682 & 0.659\\
 & \textit{comment\_type\_pharo} & 0.522 & 0.545 & 0.529 & 0.435 & 0.573 & 0.583 & 0.72 & 0.684 & 0.702 & 0.704 & 0.612 & 0.703 & 0.129 & 0.526 & 0.621 & 0.665 & 0.658 & 0.65 & 0.667 & 0.121 & 0.678 & 0.655 & 0.352 & 0.311 & 0.423 & 0.395 & 0.498 & 0.341\\
 & \textit{comment\_type\_python} & 0.574 & 0.484 & 0.638 & 0.105 & 0.471 & 0.611 & 0.64 & 0.614 & 0.708 & 0.649 & 0.633 & 0.68 & 0.422 & 0.632 & 0.596 & 0.502 & 0.602 & 0.691 & 0.667 & 0.244 & 0.665 & 0.695 & 0.383 & 0.354 & 0.373 & 0.383 & 0.524 & 0.521\\
 & \textit{review\_aspect} & 0.406 & 0.119 & 0.424 & 0.0 & 0.45 & 0.35 & 0.518 & 0.515 & 0.566 & 0.559 & 0.577 & 0.522 & 0.149 & 0.419 & 0.421 & 0.564 & 0.405 & 0.528 & 0.598 & 0.524 & 0.451 & 0.489 & 0.436 & 0.431 & 0.458 & 0.471 & 0.406 & 0.163\\
 & \textit{smell\_doc} & 0.583 & 0.155 & 0.57 & 0.517 & 0.584 & 0.639 & 0.655 & 0.648 & 0.641 & 0.703 & 0.646 & 0.616 & 0.439 & 0.516 & 0.59 & 0.611 & 0.612 & 0.622 & 0.44 & 0.609 & 0.66 & 0.625 & 0.13 & 0.398 & 0.486 & 0.473 & 0.714 & 0.614\\
Regr. & \textit{story\_points} & 0.422 & 0.283 & 0.257 & 0.254 & 0.371 & 0.379 & 0.447 & 0.434 & 0.39 & 0.409 & 0.416 & 0.46 & 0.295 & 0.371 & 0.415 & 0.429 & 0.45 & 0.464 & 0.369 & 0.361 & 0.437 & 0.464 & 0.245 & 0.214 & 0.011 & 0.043 & 0.341 & 0.253\\
NER & \textit{se\_entities} & 0.411 & 0.492 & 0.491 & 0.583 & 0.517 & 0.469 & - & - & - & - & - & - & 0.343 & 0.426 & 0.492 & 0.557 & 0.543 & - & - & - & 0.529 & 0.512 & - & - & - & - & - & -\\
MLM & \textit{requirement\_completion} & 0.5 & 0.525 & 0.626 & 0.665 & 0.695 & 0.709 & - & - & - & - & - & - & - & - & - & - & 0.624 & - & - & - & - & - & - & - & - & - & - & -\\
\begin{sideways}Family\end{sideways} & Code-adjacent metadata & M = 0.723, SD = 0.154 &  &  &  &  &  &  &  &  &  &  &  &  &  &  &  & M =~0.739, SD =~0.148 &  &  &  &  &  & M =~0.584, SD =~0.199 &  &  &  & M =~0.657 & \\
 & Core non-code & M =~0.670, SD =~0.234 &  &  &  &  &  &  &  &  &  &  &  &  &  &  &  & M =~0.694, SD =~0.196 &  &  &  &  &  & M =~0.633, SD =~0.273 &  &  &  & M =~0.527 & \\
 & Developer communication & M =~0.722, SD =~0.122 &  &  &  &  &  &  &  &  &  &  &  &  &  &  &  & M = 0.694, SD = 0.131 &  &  &  &  &  & M = 0.527, SD = 0.155 &  &  &  & M =~0.623 & \\
\begin{sideways}Task type\end{sideways} & Binary classification & M = 0.806, SD = 0.11 &  &  &  &  &  &  &  &  &  &  &  &  &  &  &  & M = 0.783, SD = 0.14 &  &  &  &  &  & M = 0.742, SD = 0.137 &  &  &  & M = 0.694 & \\
 & Multi-class classification & M = 0.75, SD = 0.124 &  &  &  &  &  &  &  &  &  &  &  &  &  &  &  & M = 0.761, SD = 0.098 &  &  &  &  &  & M = 0.629, SD = 0.158 &  &  &  & M = 0.639 & \\
 & Multi-label classification & M = 0.575, SD = 0.173 &  &  &  &  &  &  &  &  &  &  &  &  &  &  &  & M = 0.611, SD = 0.16 &  &  &  &  &  & M = 0.392, SD = 0.081 &  &  &  & M = 0.512 & \\
 & Regression & M = 0.377, SD = 0.068 &  &  &  &  &  &  &  &  &  &  &  &  &  &  &  & M = 0.424, SD = 0.047 &  &  &  &  &  & M = 0.129, SD = 0.118 &  &  &  & M = 0.297 & \\
 & NER & M = 0.478, SD = 0.071 &  &  &  &  &  &  &  &  &  &  &  &  &  &  &  & M = 0.528, SD = 0.016 &  &  &  &  &  & - &  &  &  & - & \\
 & MLM & M = 0.62, SD = 0.088 &  &  &  &  &  &  &  &  &  &  &  &  &  &  &  & M = 0.624 &  &  &  &  &  & - &  &  &  & - & 
\end{tblr}
}
\end{sidewaystable*}

\end{document}